\documentclass[12pt]{article}
\usepackage{amsmath}

\usepackage{amssymb}
\usepackage{bm}
\usepackage{braket}
\usepackage[dvips]{graphicx}

\begin{document}

\baselineskip=17pt

\begin{titlepage}
\rightline\today
\begin{center}
\vskip 2.5cm
{\Large \bf {Closed string field theory}}\\
\vskip 0.3cm
{\Large \bf {without the level-matching condition}}
\vskip 1.0cm
{\large Yuji Okawa and Ryosuke Sakaguchi}
\vskip 1.0cm
{\it {Institute of Physics, The University of Tokyo}}\\
{\it {3-8-1 Komaba, Meguro-ku, Tokyo 153-8902, Japan}}\\

\vskip 2.0cm

{\bf Abstract}
\end{center}

\noindent
In formulating covariant closed string field theories,
we have always used closed string fields with the level-matching condition.
Recently, open superstring field theories including the Ramond sector
were constructed,
and one approach was to use open superstring fields
with a constraint analogous to the level-matching condition
for closed string fields.
There is another approach developed by Sen,
where no constraints are imposed on open superstring fields
but additional free string fields are introduced
so that covariant kinetic terms for the Ramond sector were constructed.
Motivated by this development,
we construct closed string field theory
without imposing the level-matching condition on the closed string field,
while we introduce an additional free string field.
This is the first implementation of general covariance
in the context of string theory without using the level-matching condition.
We explicitly expand the string fields
up to the level relevant to massless fields,
and we confirm the equivalence to the theory
with the level-matching condition up to decoupled free fields.

\end{titlepage}

\tableofcontents


\section{Introduction}\label{Introduction}
\setcounter{equation}{0}

When we parameterize the world-sheet for free propagation of a closed string
using the coordinates $( \tau, \sigma )$ with $\sigma$ in the range $0 \le \sigma \le 2 \pi$,
we do not have an appropriate gauge-fixing condition to determine
the origin of the $\sigma$ coordinate along the closed string.
Even in the light-cone gauge, the gauge redundancy is not completely fixed
for the closed string because of this issue,
but the range of the $\sigma$ coordinate is compact
so that we can simply path integrate over all possible choices for $\sigma = 0$.
Using the zero modes $L_0$ and $\widetilde{L}_0$
of the holomorphic and antiholomorphic components, respectively,
of the energy-momentum tensor,
the translation of the $\sigma$ coordinate is generated
by $L_0 -\widetilde{L}_0$,
and the integration over all choices for $\sigma = 0$
requires the closed string state to be annihilated by $L_0 -\widetilde{L}_0$.
This is the origin of the level-matching condition in the closed string.
In closed string field theory, this is reflected
in the constraints
\begin{equation}
( L_0 -\widetilde{L}_0 ) \, \Psi = 0 \,, \qquad
( b_0 -\widetilde{b}_0 ) \, \Psi = 0
\label{level-matching-constraints}
\end{equation}
imposed on the closed string field $\Psi$,
where $b_0$ and $\widetilde{b}_0$ are the zero modes
of the $b$ ghost and the $\widetilde{b}$ ghost, respectively.
As can be understood from the origin of the level-matching condition,
it has been difficult to formulate covariant closed string field theory
without imposing these constraints on the closed string field.

In the context of the moduli space of Riemann surfaces,
the world-sheet for free propagation of a closed string has two moduli.
The moduli space can be parameterized
as $e^{t \, ( L_0 +\widetilde{L}_0 ) +i \, \theta ( L_0 -\widetilde{L}_0 )}$,
where $t$ and $\theta$ are the moduli.
In closed string field theory, the integration over $t$ is implemented by the propagator.
In Siegel gauge, the propagator can be represented as
\begin{equation}
( b_0 +\widetilde{b}_0 ) \int_0^\infty dt \, e^{t \, ( L_0 +\widetilde{L}_0 )} \,,
\end{equation}
which takes the form of an integral over the modulus $t$.
On the other hand, the integration over $\theta$ yields the operator $B$ given by
\begin{equation}
B = ( b_0 -\widetilde{b}_0 ) \int_0^{2 \pi} \frac{d \theta}{2 \pi} \,
e^{i \theta ( L_0 -\widetilde{L}_0 )} \,,
\label{B-introduction}
\end{equation}
where $b_0 -\widetilde{b}_0$ is the ghost insertion associated with the integration over this modulus.
The operator $B$ can be schematically understood
as $\delta ( b_0 -\widetilde{b}_0 ) \, \delta ( L_0 -\widetilde{L}_0 ) \,$,
and the constraints~\eqref{level-matching-constraints} on the closed string field
implement the integration over this modulus in closed string field theory.
The appropriate inner product of $\Psi_1$ and $\Psi_2$
satisfying the constraints can be written as
\begin{equation}
\langle \, \Psi_1, c_0^- \Psi_2 \, \rangle
\end{equation}
with
\begin{equation}
c_0^- = \frac{1}{2} \, ( \, c_0 -\widetilde{c}_0 \, ) \,,
\end{equation}
where $c_0$ and $\widetilde{c}_0$
are the zero modes
of the $c$ ghost and the $\widetilde{c}$ ghost, respectively,
and $\langle \, A, B \, \rangle$ is the BPZ inner product
for a pair of states $A$ and $B$.
The kinetic term for closed bosonic string field theory is then given by
\begin{equation}
S = {}-\frac{1}{2} \, \langle \, \Psi, c_0^- Q_B \Psi \, \rangle \,,
\label{closed-string-field-theory-kinetic-term}
\end{equation}
where $Q_B$ is the BRST operator.
The operator $B$ can also be written as
\begin{equation}
B = -i \int_0^{2 \pi} \frac{d \theta}{2 \pi} \int d \widetilde{\theta} \,
e^{i \theta ( L_0 -\widetilde{L}_0 ) +i \widetilde{\theta} \, ( b_0 -\widetilde{b}_0 )} \,,
\label{B}
\end{equation}
where $\widetilde{\theta}$ is a Grassmann-odd variable,
and the extended BRST transformation introduced in~\cite{Witten:2012bh}
maps $\theta$ to $\widetilde{\theta}$.
The extended BRST transformation acts in the same way
as the ordinary BRST transformation for operators
in the conformal field theory,
and in particular it maps $b_0 -\widetilde{b}_0$ to $L_0 -\widetilde{L}_0$.
Therefore, the combination $i \theta ( L_0 -\widetilde{L}_0 ) +i \widetilde{\theta} \, ( b_0 -\widetilde{b}_0 )$
in~\eqref{B} is obtained from $i \theta \, ( b_0 -\widetilde{b}_0 )$
by the extended BRST transformation.
Note that the closed bosonic string field $\Psi$ satisfying the constraints
can be characterized as
\begin{equation}
B \, c_0^- \, \Psi = \Psi \,.
\label{closed-bosoinc-string-field-charaicterized}
\end{equation}

Recently, there have been important developments
in the treatment of the Ramond sector in superstring field theory,
which we consider is related to formulating
closed string field theory without imposing the level-matching condition.
In~\cite{Kunitomo:2015usa, Erler:2016ybs, Konopka:2016grr}
complete actions for open superstring field theory
including both the Neveu-Schwarz sector and the Ramond sector
were constructed,
where the string field in the Ramond sector
is restricted to an appropriate subspace of the Hilbert space.
This constraint on the string field $\Psi$ of picture $-1/2$
can be characterized as
\begin{equation}
X Y \Psi = \Psi
\label{Ramond-constraint}
\end{equation}
with
\begin{equation}
X = \int d \zeta \int d \widetilde{\zeta} \, e^{\, \zeta G_0 -\widetilde{\zeta} \, \beta_0}
\label{X}
\end{equation}
and
\begin{equation}
Y = c_0 \, \int d \sigma \, \sigma \, e^{\sigma \gamma_0}
= {}-c_0 \, \delta' ( \gamma_0 ) \,,
\label{inversePCO}
\end{equation}
where $G_0$ is the zero mode of the supercurrent,
$\beta_0$ is the zero mode of the $\beta$ ghost,
and $\gamma_0$ is the zero mode of the $\gamma$ ghost.
The integration variable $\zeta$ is Grassmann odd
and the integration variables
$\widetilde{\zeta}$ and $\sigma$ are Grassmann even.
The integration over these Grassmann-even variables should be understood
as an algebraic operation analogous to the integration over Grassmann-odd variables~\cite{Witten:2012bh}.
The extended BRST transformation introduced in~\cite{Witten:2012bh}
maps $\zeta$ to $\widetilde{\zeta}$
and maps $\beta_0$ to $G_0$
so that the combination $\zeta G_0 -\widetilde{\zeta} \, \beta_0$
in~\eqref{X} is obtained from $-\zeta \beta_0$
by the extended BRST transformation.

In the context of the supermoduli space of super-Riemann surfaces,
this constraint can be understood in the following way.
Propagator strips for the Ramond sector of the open superstring
have a fermionic modulus in addition to the bosonic modulus
corresponding to the length of the strip.
The fermionic direction of the moduli space can be parameterized as
$e^{\, \zeta G_0}$,
where $\zeta$ is the fermionic modulus.
The integration over $\zeta$ with the associated ghost insertion
yields the operator $X$ in~\eqref{X}.
The constraint~\eqref{Ramond-constraint}
is therefore analogous to~\eqref{closed-bosoinc-string-field-charaicterized}
for the closed bosonic string field.
The appropriate inner product of $\Psi_1$ and $\Psi_2$ in the restricted space
can be written as
\begin{equation}
\langle \, \Psi_1, Y \Psi_2 \, \rangle \,,
\end{equation}
and the kinetic term of open superstring field theory for the Ramond sector is 
given by
\begin{equation}
S = {}-\frac{1}{2} \, \langle \, \Psi, Y Q_B \Psi \, \rangle \,.
\end{equation}
This is analogous to the kinetic term~\eqref{closed-string-field-theory-kinetic-term}
for closed bosonic string field theory.

Another remarkable development in the treatment of the Ramond sector
in superstring field theory
is the construction of covariant kinetic terms by Sen~\cite{Sen:2015uaa},
where no constraints associated with the Ramond sector are imposed
but spurious free fields are introduced.
The construction was presented in the context of the Batalin-Vilkovisky master action
for heterotic string field theory or type II superstring field theory
in~\cite{Sen:2015uaa},
but the idea can be applied to the construction of a classical gauge-invariant action
for open superstring field theory.
In this context, the kinetic terms are given by
\begin{equation}
S = \frac{1}{2} \, \langle \, \widetilde{\Psi}, Q_B X \widetilde{\Psi} \, \rangle
+\langle \, \widetilde{\Psi}, Q_B \Psi \, \rangle \,,
\end{equation}
where $\Psi$ is the string field of picture $-1/2$
and $\widetilde{\Psi}$ is the string field of picture $-3/2$.
The interaction terms do not contain $\widetilde{\Psi}$
and the string field $\widetilde{\Psi}$ describes the spurious free fields.
The string field $\Psi$ describes the interacting fields
and no constraints are imposed on $\Psi$.
In this approach the operator $X$ can be replaced
by a different operator.
In~\cite{Sen:2015uaa} the zero mode of the picture-changing operator was used,
which is convenient when we describe the superconformal ghost sector
in terms of $\xi$, $\eta$, and $\phi$
introduced in~\cite{Friedan:1985ge}.

This construction indicates that we can formulate closed string field theory
without imposing the level-matching condition on the closed string field
if we allow spurious free fields.
In this paper we claim that this is indeed possible.
Our kinetic terms for closed bosonic string field theory
without the level-matching condition are given by
\begin{equation}
S = \frac{1}{2} \, \langle \, \widetilde{\Psi}, Q_B B \widetilde{\Psi} \, \rangle
+\langle \, \widetilde{\Psi}, Q_B \Psi \, \rangle \,,
\end{equation}
where $\Psi$ is the string field of ghost number $2$
and $\widetilde{\Psi}$ is the string field of ghost number $3$.
The interaction terms do not contain $\widetilde{\Psi}$
and the string field $\widetilde{\Psi}$ describes the spurious free fields.
The string field $\Psi$ describes the interacting fields
and no constraints are imposed on $\Psi$.

The rest of the paper is organized as follows.
In section~\ref{Review} we present closed bosonic string field theory
without imposing the level-matching condition on the closed string field.
We briefly review closed bosonic string field theory
with the level-matching condition in subsection~\ref{section-2.1},
and then we present the action and the gauge transformation
of closed bosonic string field theory
without the level-matching condition in subsection~\ref{section-2.2}.
This is the main result of this paper.
In section~\ref{expansion} we expand string fields in terms of component fields
up to the level relevant to massless fields,
and we demonstrate the equivalence of
the theory without the level-matching condition
to the theory with the level-matching condition
up to extra free fields.\footnote{
Component fields are studied in more detail
by Erbin and Med\'{e}vielle~\cite{Erbin-Medevielle}.
}
Section~\ref{conclusions-discussion} is devoted to conclusions and discussion.

\section{Action}\label{Review}
\setcounter{equation}{0}

\subsection{Closed string field theory with the level-matching condition}
\label{section-2.1}
In closed bosonic string field theory, whose construction \cite{Kaku:1988zv,Kaku:1988zw,Saadi:1989tb,Kugo:1989aa,Kugo:1989tk} was completed by Zwiebach in \cite{Zwiebach:1992ie},
we use the closed string field $A$ which satisfies the constraints
\begin{equation}
L_{0}^{-} A = 0
\label{b}
\end{equation}
and
\begin{equation}
b_{0}^{-} A = 0 \,,
\label{bb}
\end{equation}
where
\begin{equation}
L_{0}^{-} = L_{0}-\widetilde{L}_{0} \,, \qquad
b_{0}^{-} = b_{0}-\widetilde{b}_{0} \,.
\end{equation}
The string field $A$ satisfying the constraints $(\ref{b})$ and $(\ref{bb})$ can also be characterized as
 \begin{eqnarray}
 \begin{split}
Bc_{0}^{-} A = A \,,
\label{bc}
 \end{split}
\end{eqnarray}
where $B$ and $c_{0}^{-}$ are defined by
\begin{equation}
B =b_{0}^{-}\int_{0}^{2\pi}\frac{d\theta}{2\pi } \, e^{i\theta L_{0}^{-}} \,, \qquad
c_{0}^{-} = \frac{1}{2}\, (c_{0}-\widetilde{c}_{0}) \,.
\end{equation}
The operator $B$ anticommutes with the BRST operator $Q_{B}$, 
\begin{equation}
\{ Q_{B},B\} =0 \,,
\end{equation}
and $B$ is BPZ even:
 \begin{eqnarray}
 \begin{split}
\braket{BA_{1},A_{2}}&=(-1)^{A_{1}}\braket{A_{1},BA_{2}}.&
 \end{split}
\end{eqnarray}
Here and in what follows, a state in the exponent of $-1$ represents its Grassmann parity: it is $0$ mod $2$ for a Grassmann-even state and $1$ mod $2$ for a Grassmann-odd state.
Since
\begin{equation}
B c_0^- B = B \,,
\end{equation}
the operator $B c_0^-$ is a projector
onto a subspace of states annihilated by $L_0^-$ and $b_0^-$.

The kinetic term of the closed string field $\Psi$ of ghost number $2$ is given by
\begin{equation}
\frac{1}{2}\langle \! \langle \Psi ,Q_{B}\Psi \rangle \! \rangle \,,
\end{equation}
where $\langle \! \langle A_{1},A_{2} \rangle \! \rangle$ is defined by
\begin{eqnarray}
 \begin{split}
\langle \! \langle A_{1},A_{2} \rangle \! \rangle =\braket{A_{1},c_{0}^{-}A_{2}}\,.
 \end{split}
\end{eqnarray}
The inner product $\langle \! \langle A_{1},A_{2} \rangle \! \rangle$
for states $A_{1}$ and $A_{2}$ satisfying the constraints \eqref{b} and \eqref{bb}
has the following properites:
 \begin{eqnarray}
 \begin{split}
\langle \! \langle A_{1},A_{2}\rangle \! \rangle &=(-1)^{(A_{1}+1)(A_{2}+1)}\langle \! \langle  A_{2},A_{1}\rangle \! \rangle , &\\
\langle \! \langle Q_{B}A_{1},A_{2} \rangle \! \rangle &=(-1)^{A_{1}} \langle \! \langle A_{1},Q_{B}A_{2} \rangle \! \rangle ,&  \label{aaa}
\end{split}
\end{eqnarray}
which can be shown from
\begin{eqnarray}
 \begin{split}
\braket{A_{1},A_{2}}&=(-1)^{A_{1}A_{2}}\braket{A_{2},A_{1}}, &\\
\braket{Q_{B}A_{1},A_{2}}&=-(-1)^{A_{1}}\braket{A_{1},Q_{B}A_{2}}.&
 \end{split}
\end{eqnarray}

The action including interactions is given by
 \begin{eqnarray}
  \begin{split}
S&=\frac{1}{2}\langle \! \langle \Psi ,Q_{B}\Psi \rangle \! \rangle +\sum_{n=3}^{\infty}\frac{g^{n-2}}{n!}\langle \! \langle \Psi ,[\![\, \underbrace{ \Psi ,\Psi ,\cdots ,\Psi }_{n-1} \, ]\!]  \rangle \! \rangle &\\
&=\frac{1}{2}\langle \! \langle \Psi ,Q_{B}\Psi \rangle \! \rangle +\frac{g}{3!}\langle \! \langle \Psi ,[\![ \Psi ,\Psi ]\!] \rangle \! \rangle +\frac{g^{2}}{4!}\langle \! \langle \Psi ,[\![ \Psi ,\Psi ,\Psi ]\!] \rangle \! \rangle +O(g^{3}),& \label{a}
 \end{split}
\end{eqnarray}
where $g$ is the closed string coupling constant
and the $n$-string product $[\![  A_{1},A_{2},\cdots ,A_{n}]\!]$
is defined for states $A_1, A_2, \cdots , A_n$
satisfying the constraints \eqref{b} and \eqref{bb}.
It takes the form
 \begin{equation}
  \begin{split}
[\![  A_{1},A_{2},\cdots ,A_{n}]\!] = B \, \mathcal{F} \, ( \, A_{1},A_{2},\cdots ,A_{n} \, ) \,, \label{2000}
 \end{split}
\end{equation}
where the operator $B$ multiplies
an $n$-string product $\mathcal{F} \, ( \, A_{1},A_{2},\cdots ,A_{n} \, )$,
so that the $n$-string product $[\![  A_{1},A_{2},\cdots ,A_{n}]\!]$
satisfies the constraints \eqref{b} and \eqref{bb}.
The ghost number of the $n$-string product $G([\![ A_{1},A_{2},\cdots ,A_{n}]\!] )$ is given by
\begin{eqnarray}
G([\![ A_{1},A_{2},\cdots ,A_{n}]\!] )=-2(n-2)-1+\sum_{i=1}^{n}G(A_{i}),
\end{eqnarray}
where $G(A)$ is the ghost number of $A$.
The $n$-string product is graded-commutative,
\begin{eqnarray}
[\![ A_{1},\cdots ,A_{i+1},A_{i},A_{i+2},\cdots ,A_{n}]\!]  =(-1)^{A_{i}A_{i+1}}[\![ A_{1},\cdots ,A_{n}]\!] \,,
\end{eqnarray}
and the inner product $\langle \! \langle A_{1}, [\![ A_{2},\cdots ,A_{n}]\!] \rangle \! \rangle $ has the following property:
\begin{eqnarray}
\langle \! \langle A_{1}, [\![ A_{2},\cdots ,A_{n}]\!] \rangle \! \rangle =(-1)^{A_{1}A_{2}}\langle \! \langle A_{2}, [\![ A_{1},\cdots ,A_{n}]\!] \rangle \! \rangle .
\end{eqnarray}
Therefore, the multi-linear function defined by
\begin{eqnarray}
\{ \! \! \{  A_{1},A_{2},\cdots ,A_{n}\} \! \! \} \equiv \langle \! \langle A_{1},[\![ A_{2},\cdots ,A_{n}]\!]\rangle \! \rangle
\end{eqnarray}
is graded-commutative:
\begin{eqnarray}
\{ \! \! \{ A_{1},\cdots ,A_{i+1},A_{i},A_{i+2},\cdots ,A_{n}\} \! \! \} =(-1)^{A_{i}A_{i+1}}\{ \! \! \{ A_{1},\cdots ,A_{n}\} \! \! \} .
\end{eqnarray}
We use the notation $\Psi ^{n}$
for $n$ successive entries of the same string field $\Psi$
in the multi-string products or the multi-linear functions:
  \begin{eqnarray}
 \begin{split}
\Psi ^{n} \equiv \underbrace{ \Psi ,\Psi ,\cdots ,\Psi }_{n}.
 \end{split}
\end{eqnarray}
With this notation, the action can be expressed as
 \begin{eqnarray}
  \begin{split}
S&=\frac{1}{2}\langle \! \langle \Psi ,Q_{B}\Psi \rangle \! \rangle +\sum_{n=3}^{\infty}\frac{g^{n-2}}{n!} \{ \! \! \{   \Psi ^{n}  \} \! \! \} \,. &
 \end{split}
\end{eqnarray}

The variation of the action is given by
 \begin{eqnarray}
  \begin{split}
\delta S&=\langle \! \langle \delta \Psi ,Q_{B}\Psi\rangle \! \rangle  +\sum_{n=2}^{\infty}\frac{g^{n-1}}{n!}\langle \! \langle \delta \Psi ,[\![ \Psi ^{n} ]\!] \rangle \! \rangle  &\\
&=\langle \! \langle  \delta \Psi ,Q_{B}\Psi \rangle \! \rangle +\frac{g}{2!}\langle \! \langle \delta \Psi ,[\![ \Psi ,\Psi ]\!]\rangle \! \rangle +\frac{g^{2}}{3!}\langle \! \langle \delta \Psi ,[\![ \Psi ,\Psi ,\Psi ]\!]\rangle \! \rangle +O(g^{3}).& \\ \label{k}
 \end{split}
\end{eqnarray}
The equation of motion is then
 \begin{eqnarray}
Q_{B}\Psi+\sum_{n=2}^{\infty}\frac{g^{n-1}}{n!}[\![ \Psi ^{n}]\!] =0. \label{c}
\end{eqnarray} 
The action $(\ref{a})$ is invariant under the gauge transformation
\begin{eqnarray}
\begin{split}
\delta _{\Lambda}\Psi &=Q_{B}\Lambda +\sum_{n=1}^{\infty}\frac{g^{n}}{n!}[\![ \Psi ^{n},\Lambda ]\!] &\\
&=Q_{B}\Lambda +g[\![ \Psi ,\Lambda ]\!] +\frac{g^{2}}{2!}[\![ \Psi ,\Psi ,\Lambda ]\!] +O(g^{3})& \label{i}
\end{split}
\end{eqnarray}
for the gauge parameter $\Lambda$ satisfying 
 \begin{eqnarray}
 \begin{split}
L_{0}^{-}\Lambda =0,\\
b_{0}^{-}\Lambda =0, 
 \end{split}
\end{eqnarray}
if the multi-string products satisfy 
\begin{eqnarray}
\begin{split}
0=&Q_{B}[\![ A_{1},\cdots ,A_{n}]\!] +\sum_{i=1}^{n}(-1)^{(A_{1}+\cdots +A_{i-1})}[\![ A_{1},\cdots ,Q_{B}A_{i},\cdots ,A_{n}]\!] &\\
&+\sum_{\substack{\{ i_{l},j_{k}\}\\ l\geq 1,k\geq 2 }}\sigma ( i_{l},j_{k})[\![ A_{i_{1}},\cdots ,A_{i_{l}},[\![ A_{j_{1}}\cdots ,A_{j_{k}}]\!] ]\!].& \label{g}
\end{split}
\end{eqnarray}
The second summation on the right-hand side runs over all different splittings of the set $\{ 1,\cdots , n\}$ into a first group $\{ i_{1},\cdots , i_{l}\}$ and a second group $\{ j_{1},\cdots , j_{k}\} $. The sign factor $\sigma ( i_{l},j_{k})$ is the sign picked up when we rearrange the sequence $\{ Q_{B},A_{1},\cdots ,A_{n}\} $ into the sequence $\{ A_{i_{1}},\cdots ,A_{i_{l}},Q_{B},A_{j_{1}},\cdots ,A_{j_{k}}\}$ taking into account the Grassmann property of the various objects. 
The relations~\eqref{g} among multi-string products
are called $L_\infty$ relations.

Let us see the invariance of the action up to $O(g^{3})$ under the transformation $(\ref{i})$ explicitly. We expand the action $(\ref{a})$ and the gauge transformation $(\ref{i})$ as follows:
 \begin{eqnarray}
  \begin{split}
S&=S^{(0)}+g\, S^{(1)}+g^{2}\, S^{(2)}+O(g^{3}),& \\
\delta _{\Lambda}\Psi &=\delta _{\Lambda}^{(0)}\Psi +g\, \delta _{\Lambda}^{(1)}\Psi +g^{2}\, \delta _{\Lambda}^{(2)}\Psi +O(g^{3}),&
 \end{split}
\end{eqnarray}
where 
 \begin{eqnarray}
  \begin{split}
&S^{(0)}=\frac{1}{2} \langle \! \langle \Psi ,Q_{B}\Psi \rangle \! \rangle ,\ \ \  S^{(1)}=\frac{1}{3!}\langle \! \langle \Psi ,[\![ \Psi ,\Psi ]\!] \rangle \! \rangle,\ \ \ S^{(2)}=\frac{1}{4!}\langle \! \langle \Psi ,[\![ \Psi ,\Psi ,\Psi ]\!] \rangle \! \rangle, &\\
&\delta _{\Lambda}^{(0)}\Psi =Q_{B}\Lambda ,\ \ \ \ \ \ \ \ \ \ \ \ \ \  \delta _{\Lambda}^{(1)}\Psi =[\![ \Psi ,\Lambda ]\!] ,\ \ \ \ \ \ \ \ \ \ \ \ \ \ \ \delta _{\Lambda}^{(2)}\Psi =\frac{1}{2!}[\![ \Psi ,\Psi ,\Lambda ]\!] .&
 \end{split}
\end{eqnarray}
The variation $\delta _{\Lambda}^{(0)}S^{(0)}$ is
 \begin{eqnarray}
  \begin{split}
\delta _{\Lambda}^{(0)}S^{(0)}=\langle \! \langle Q_{B}\Lambda ,Q_{B}\Psi\rangle \! \rangle .
\label{q}
\end{split}
\end{eqnarray}
The gauge invariance at this order follows from the nilpotency of $Q_{B}$. 
The variation at $O(g)$ is
 \begin{eqnarray}
  \begin{split}
\delta _{\Lambda}^{(0)}S^{(1)}+\delta _{\Lambda}^{(1)}S^{(0)}&=\frac{1}{2!}\langle \! \langle Q_{B}\Lambda ,[\![ \Psi ,\Psi ]\!] \rangle \! \rangle +\langle \! \langle [\![ \Psi ,\Lambda ]\!] ,Q_{B}\Psi \rangle \! \rangle&\\
&=-\frac{1}{2!}\langle \! \langle \Lambda ,Q_{B}[\![ \Psi ,\Psi ]\!] \rangle \! \rangle  -\langle \! \langle  \Lambda ,[\![ Q_{B}\Psi ,\Psi ]\!] \rangle \! \rangle  .&
\label{r}
\end{split}
\end{eqnarray}
The gauge invariance at this order follows from
\begin{eqnarray}
\begin{split}
0=Q_{B}[\![ A_{1},A_{2}]\!] +[\![ Q_{B}A_{1},A_{2}]\!] +(-1)^{A_{1}}[\![ A_{1},Q_{B}A_{2}]\!] ,
\end{split}
\end{eqnarray}
which is the relation~\eqref{g} for $n=2$. 
The variation at $O(g^{2})$ is 
 \begin{eqnarray}
  \begin{split}
 & \delta _{\Lambda}^{(0)}S^{(2)}+\delta _{\Lambda}^{(1)}S^{(1)}+\delta _{\Lambda}^{(2)}S^{(0)}&\\
  =&\frac{1}{3!}\langle \! \langle Q_{B}\Lambda  ,[\![ \Psi ,\Psi ,\Psi ]\!] \rangle \! \rangle +\frac{1}{2!}\langle \! \langle [\![ \Psi ,\Psi ,\Lambda ]\!] ,Q_{B}\Psi \rangle \! \rangle +\frac{1}{2!}\langle \! \langle [\![ \Psi ,\Lambda ]\!] ,[\![ \Psi ,\Psi ]\!] \rangle \! \rangle & \\
=&-\frac{1}{3!}\langle \! \langle  \Lambda  ,Q_{B}[\![ \Psi ,\Psi ,\Psi ]\!] \rangle \! \rangle -\frac{1}{2!}\langle \! \langle \Lambda ,[\![ Q_{B}\Psi ,\Psi ,\Psi ]\!] \rangle \! \rangle -\frac{1}{2!}\langle \! \langle \Lambda ,[\![ \Psi ,[\![ \Psi ,\Psi ]\!] ]\!] \rangle \! \rangle .&
\label{s}
\end{split}
\end{eqnarray}
The gauge invariance at this order follows from
\begin{eqnarray}
0=Q_B [\![ A_{1},A_{2},A_{3}]\!] +[\![ Q_{B}A_{1},A_{2},A_{3}]\!]+(-1)^{A_{1}}[\![ A_{1},Q_{B}A_{2},A_{3}]\!] +(-1)^{A_{1}+A_{2}}[\![ A_{1},A_{2},Q_{B}A_{3}]\!]  \nonumber \\
+(-1)^{A_{1}}[\![ A_{1},[\![ A_{2},A_{3}]\!] ]\!]+(-1)^{A_{2}(1+A_{1})}[\![ A_{2},[\![ A_{1},A_{3}]\!] ]\!]+(-1)^{A_{3}(1+A_{1}+A_{2})}[\![ A_{3},[\![ A_{1},A_{2}]\!] ]\!], \nonumber \\
\end{eqnarray}
which is the relation~\eqref{g} for $n=3$.

\subsection{Closed string field theory without the level-matching condition}
\label{section-2.2}
Let us move on to the construction
of closed string field theory without imposing the level-matching condition.
We use two string fields $\Psi$ and $\widetilde{\Psi}$ carrying ghost number $2$ and $3$, respectively,
where the constraints~\eqref{b} and~\eqref{bb} are not imposed on these string fields.
As we discussed in the introduction,
let us consider the kinetic terms given by
\begin{equation}
\frac{1}{2}\braket{\widetilde{\Psi},Q_{B}B\widetilde{\Psi}}
+\braket{\widetilde{\Psi},Q_{B}\Psi} \,,
\end{equation}
which are invariant under the following gauge transformations
with gauge parameters $\Lambda$ and $\widetilde{\Lambda}$:
\begin{align}
\delta _{\Lambda}\Psi &=Q_{B}\Lambda \,, &
\delta _{\widetilde{\Lambda}}\Psi  &=0 \,, & \\
\delta _{\Lambda}\widetilde{\Psi} &= 0 \,, &
\delta _{\widetilde{\Lambda}}\widetilde{\Psi}& =Q_{B}\widetilde{\Lambda} \,. &
\end{align}
The equations of motion derived from the kinetic terms are
\begin{align}
Q_{B}B\widetilde{\Psi}+Q_{B}\Psi & = 0 \,, \\
Q_{B}\widetilde{\Psi} & = 0 \,.
\end{align}
Since $B$ anticommutes with the BRST operator,
we can eliminate $\widetilde{\Psi}$ from the first equation
using $Q_{B}\widetilde{\Psi} = 0$ to find
\begin{equation}
Q_{B}\Psi = 0 \,.
\end{equation}
While we do not impose the constraints~\eqref{b} and~\eqref{bb} on $\Psi$,
it is known that the BRST cohomology on the space of states
without these constraints
is the same as that on the space of states
with these constraints.\footnote{
We will not discuss possible subtleties associated with zero-momentum states.
}
Therefore, the correct spectrum is reproduced from the kinetic terms
up to additional physical states from $\widetilde{\Psi}$.

The next step is to include interactions so that the resulting action
is invariant under nonlinearly extended gauge transformations.
In the approach by Sen,
the interaction terms in the equation of motion
for the Ramond sector are written in terms of
multi-string products with a factor of the operator $X$
and the associated terms in the action are constructed from
multi-string products without the factor of $X$.
In the construction of closed string field theory
reviewed in the preceding subsection,
the interaction terms in the equation of motion
are written in terms of
multi-string products with a factor of the operator $B$
as we mentioned in~\eqref{2000}:
\begin{equation}
[\![  A_{1},A_{2},\cdots ,A_{n}]\!]
= B \, \mathcal{F} ( \, A_{1},A_{2},\cdots ,A_{n} \, )
= b_0^- \int_0^{2 \pi} \frac{d \theta}{2 \pi} \,
e^{i \theta L_0^-} \,
\mathcal{F} ( \, A_{1},A_{2},\cdots ,A_{n} \, ) \,.
\label{double-bracket-from-F}
\end{equation}
Based on the analogy with the approach by Sen,
we want to use multi-string products without the factor of $B$,
but it seems difficult to construct consistent interactions
without the integration over $\theta$.
It also seems difficult to construct consistent interactions
when $A_1$, $A_2$, $\cdots$, $A_n$ do not satisfy
the level-matching condition~\eqref{b}.
We therefore use the projector $B c_0^-$
and remove only the factor of $b_0^-$ to define
the multi-string products $[ A_{1},A_{2},\cdots ,A_{n}]$ as follows:
\begin{equation}
[ A_{1},A_{2},\cdots ,A_{n}]
= \int_0^{2 \pi} \frac{d \theta}{2 \pi} \,
e^{i \theta L_0^-} \,
\mathcal{F} ( \, B c_0^- A_{1}, B c_0^- A_{2},\cdots , B c_0^- A_{n} \, ) \,.
\label{single-bracket-from-F}
\end{equation}
Since
\begin{equation}
\begin{split}
B \int_0^{2 \pi} \frac{d \theta}{2 \pi} \,
e^{i \theta L_0^-}
= b_0^-\, \biggl( \, \int_0^{2 \pi} \frac{d \theta}{2 \pi} \,
e^{i \theta L_0^-} \, \biggr)^2
= b_0^- \, \int_0^{2 \pi} \frac{d \theta}{2 \pi} \,
e^{i \theta L_0^-}
= B \,,
\end{split}
\end{equation}
the multi-string products 
$[\![  A_{1},A_{2},\cdots ,A_{n}]\!]$ and $[ A_{1},A_{2},\cdots ,A_{n}]$ are related as
 \begin{equation}
  \begin{split}
[\![  A_{1},A_{2},\cdots ,A_{n}]\!] = B \, [ A_{1},A_{2},\cdots ,A_{n}]
\label{identification-of-string-products}
 \end{split}
\end{equation}
when all of $A_1$, $A_2$, $\cdots$, $A_n$ satisfy the constraints~\eqref{b} and~\eqref{bb}.
In this sense the multi-string products $[ A_{1},A_{2},\cdots ,A_{n}]$
can be thought of as those without the factor of $B$ in $[\![  A_{1},A_{2},\cdots ,A_{n}]\!]$. 
Let us now consider an action of the following form:
\begin{eqnarray}
\begin{split}
S&=\frac{1}{2}\braket{\widetilde{\Psi},Q_{B}B\widetilde{\Psi}}+\braket{\widetilde{\Psi},Q_{B}\Psi}+\sum_{n=3}^{\infty}\frac{g^{n-2}}{n!}\braket{ \Psi ,[ \, \Psi ^{n-1} \, ]  } &\\
&=\frac{1}{2}\braket{\widetilde{\Psi},Q_{B}B\widetilde{\Psi}}+\braket{\widetilde{\Psi},Q_{B}\Psi}+\frac{g}{3!}\braket{\Psi ,[\Psi ,\Psi ]}+\frac{g^{2}}{4!}\braket{\Psi ,[\Psi ,\Psi ,\Psi ]}+O(g^{3}) \,. & \label{d}
\end{split}
\end{eqnarray}
The ghost number of $[ A_{1},A_{2},\cdots ,A_{n}] $ is 
\begin{eqnarray}
G([ A_{1},A_{2},\cdots ,A_{n}] )=-2(n-2)+\sum_{i=1}^{n}G(A_{i}) \,.
\end{eqnarray}
The $n$-string product is graded-commutative,
\begin{eqnarray}
[ A_{1},\cdots ,A_{i+1},A_{i},A_{i+2},\cdots ,A_{n}]  =(-1)^{A_{i}A_{i+1}}[ A_{1},\cdots ,A_{n}] ,
\end{eqnarray}
and the inner product $\langle  A_{1}, [ A_{2},\cdots ,A_{n}]  \rangle $ has the following property:
\begin{eqnarray}
\langle A_{1}, [ A_{2},\cdots ,A_{n}] \rangle =(-1)^{A_{1}A_{2}}\langle A_{2}, [ A_{1},\cdots ,A_{n}] \rangle .
\end{eqnarray}
Therefore, the multi-linear function defined by
\begin{eqnarray}
\{  A_{1},A_{2},\cdots ,A_{n}\}  \equiv  \langle A_{1},[A_{2},\cdots ,A_{n}]\rangle 
\end{eqnarray}
is graded-commutative:
\begin{eqnarray}
\{  A_{1},\cdots ,A_{i+1},A_{i},A_{i+2},\cdots ,A_{n}\} =(-1)^{A_{i}A_{i+1}} \{ A_{1},\cdots ,A_{n}\} .
\end{eqnarray}

The variation of the action is given by
\begin{eqnarray}
\begin{split}
\delta S&=\braket{\delta \widetilde{\Psi},Q_{B}B\widetilde{\Psi}}+\braket{\delta \widetilde{\Psi},Q_{B}\Psi}+\braket{\delta \Psi ,Q_{B}\widetilde{\Psi}}+\sum_{n=2}^{\infty}\frac{g^{n-1}}{n!}\{ \delta \Psi ,\Psi ^{n} \} &\\
&=\braket{\delta \widetilde{\Psi},Q_{B}B\widetilde{\Psi}}+\braket{\delta \widetilde{\Psi},Q_{B}\Psi}+\braket{\delta \Psi ,Q_{B}\widetilde{\Psi}}&\\
&\ \ \ \ \ \ \ \ \ \ \ \ \ \ \ \ \ \ \ \ \ \ \ \ \ \ \ \ \ \ +\frac{g}{2!}\braket{\delta \Psi ,[\Psi ,\Psi ]}+\frac{g^{2}}{3!}\braket{\delta \Psi ,[\Psi ,\Psi ,\Psi ]}+O(g^{3}) \,.& \label{l}
\end{split}
\end{eqnarray}
The equations of motion are then
\begin{align}
Q_{B}B\widetilde{\Psi}+Q_{B}\Psi & = 0 \,,
\label{delta-Psi-equation} \\
Q_{B}\widetilde{\Psi}+\sum_{n=2}^{\infty}\frac{g^{n-1}}{n!}[ \, \Psi ^{n} \, ] & = 0 \,.
\label{Psi-tilde-equation}
\end{align}
We can eliminate $\widetilde{\Psi}$ 
by multiplying~\eqref{Psi-tilde-equation} by $B$
and adding the resulting equation to~\eqref{delta-Psi-equation}.
We then obtain
\begin{equation}
Q_{B}\Psi+\sum_{n=2}^{\infty}\frac{g^{n-1}}{n!} B \, [ \, \Psi ^{n} \, ] = 0 \,.
\label{Psi-equation} 
\end{equation}

Let us demonstrate the equivalence of this new theory
to the theory in the preceding subsection
up to additional free fields.
First, any solution to~\eqref{c} also solves~\eqref{Psi-equation}
because of the relation~\eqref{identification-of-string-products}.
Second, consider whether any solution to~\eqref{Psi-equation} can be brought
to a solution to~\eqref{c} by a gauge transformation.
We expand $\Psi$ in $g$ as follows:
\begin{equation}
\Psi = \sum_{n=0}^\infty g^n \, \Psi^{(n)}
= \Psi^{(0)} +g \, \Psi^{(1)} +g^2 \, \Psi^{(2)} \ldots .
\end{equation}
The equation of motion for $\Psi^{(0)}$ is
\begin{equation}
Q_B \Psi^{(0)} = 0 \,.
\label{linearized-Psi-equation}
\end{equation}
As we discussed before,
any solution to~\eqref{linearized-Psi-equation}
can be brought to a state satisfying the constraints~\eqref{b} and~\eqref{bb}
by a gauge transformation
because the cohomology of $Q_B$
for the space with the constraints~\eqref{b} and~\eqref{bb}
is the same as the cohomology of $Q_B$
for the space without the constraints.
The equation of motion for $\Psi^{(1)}$ is
\begin{equation}
Q_B \Psi^{(1)} = -\frac{1}{2} \, B \, [ \, \Psi^{(0)}, \Psi^{(0)} \, ] \,.
\label{Psi^(1)-equation}
\end{equation}
Since
\begin{equation}
Q_B \, B \, [ \, \Psi^{(0)}, \Psi^{(0)} \, ]
= -B \, Q_B \, [ \, \Psi^{(0)}, \Psi^{(0)} \, ]
= -B \, [ \, Q_B \, \Psi^{(0)}, \Psi^{(0)} \, ]
-B \, [ \, \Psi^{(0)}, Q_B \, \Psi^{(0)} \, ] \,,
\end{equation}
the right-hand side of~\eqref{Psi^(1)-equation} is BRST closed
when $\Psi^{(0)}$ satisfies~\eqref{linearized-Psi-equation}.
The BRST cohomology on the space of ghost number $3$ is nontrivial
so that the right-hand side of~\eqref{Psi^(1)-equation} may not be BRST exact in general.
In order to have $\Psi^{(1)}$ solving~\eqref{Psi^(1)-equation},
the right-hand side of~\eqref{Psi^(1)-equation} has to be BRST exact
and this is the case we are considering.
Once we have $\Psi^{(1)}$ solving~\eqref{Psi^(1)-equation},
we can again bring it to a form which satisfies the constraints~\eqref{b} and~\eqref{bb}
by a gauge transformation at~$O(g)$.
It is obvious that we can proceed to higher orders in $g$.
We thus conclude that any perturbative solution to~\eqref{Psi-equation}
can be brought to a solution to~\eqref{c} by a gauge transformation.

Let us next consider~\eqref{Psi-tilde-equation}.
This can be regarded as an equation for $\widetilde{\Psi}$
when $\Psi$ satisfying~\eqref{Psi-equation} is given.
We need to make sure that no additional conditions
are imposed on solutions to~\eqref{Psi-equation}
when we solve~\eqref{Psi-tilde-equation}.\footnote{
The following argument is a translation
of the discussion
in the context of the Ramond sector
by Sen explained in footnote~$7$ of~\cite{Sen:2015uaa}
into our context.
} 
We expand $\widetilde{\Psi}$ in $g$ as follows:
\begin{equation}
\widetilde{\Psi} = \sum_{n=0}^\infty g^n \, \widetilde{\Psi}^{(n)}
= \widetilde{\Psi}^{(0)} +g \, \widetilde{\Psi}^{(1)} +g^2 \, \widetilde{\Psi} \ldots .
\end{equation}
The equation of motion for $\widetilde{\Psi}^{(0)}$ is
\begin{equation}
Q_B \widetilde{\Psi}^{(0)} = 0 \,.
\end{equation}
The solution to this equation describes the spurious free fields in the free theory.
The equation of motion for $\widetilde{\Psi}^{(1)}$ is
\begin{equation}
Q_B \widetilde{\Psi}^{(1)} = -\frac{1}{2} \, [ \, \Psi^{(0)}, \Psi^{(0)} \, ] \,.
\label{Psi-tilde^(1)-equation}
\end{equation}
Since
\begin{equation}
Q_B \, [ \, \Psi^{(0)}, \Psi^{(0)} \, ]
= [ \, Q_B \, \Psi^{(0)}, \Psi^{(0)} \, ]
+[ \, \Psi^{(0)}, Q_B \, \Psi^{(0)} \, ] \,,
\end{equation}
the right-hand side of~\eqref{Psi-tilde^(1)-equation} is BRST closed
when $\Psi^{(0)}$ satisfies~\eqref{linearized-Psi-equation}.
Since the BRST cohomology on the space of ghost number $4$ is nontrivial,
the right-hand side of~\eqref{Psi-tilde^(1)-equation} may not be BRST exact.
If this is the case, 
we do not have $\widetilde{\Psi}^{(1)}$ solving~\eqref{Psi-tilde^(1)-equation}
and additional conditions are imposed on solutions to~\eqref{Psi-equation}.
It is known that any BRST-closed state of ghost number $4$
can be written as
\begin{equation}
c_0 \widetilde{c}_0 c_1\widetilde{c}_1 | \psi \rangle 
\label{cohomology-ghost-number-4}
\end{equation}
up to a BRST-exact piece,
where $| \psi \rangle$ is a state
corresponding to a primary field of weight $(1,1)$
in the matter sector,
and we expand the $c$ ghost and the $\widetilde{c}$ ghost as follows:
\begin{equation}
c (z) = \sum_{n=-\infty}^\infty \frac {c_n}{z^{n-1}} \,, \qquad
\widetilde{c} (\bar{z}) = \sum_{n=-\infty}^\infty
\frac {\widetilde{c}_n}{\bar{z}^{\, n-1}} \,.
\end{equation}
Suppose that $[ \, \Psi^{(0)}, \Psi^{(0)} \, ]$ contains a state of this form.
Then $B \, [ \, \Psi^{(0)}, \Psi^{(0)} \, ]$ contains a state of the form
\begin{equation}
( c_0 +\widetilde{c}_0 ) \, c_1\widetilde{c}_1 | \psi \rangle
\label{cohomology-ghost-number-3}
\end{equation}
up to a BRST-exact piece,
and it is a state in the nontrivial BRST cohomology of ghost number~$3$.
In this case we do not have $\Psi^{(1)}$ solving~\eqref{Psi^(1)-equation}.
Conversely, absence of terms of the form~\eqref{cohomology-ghost-number-3}
at ghost number $3$
implies absence of terms of the form~\eqref{cohomology-ghost-number-4}
before acting $B$ at ghost number $4$.
Therefore, when we have a solution $\Psi^{(1)}$ to~\eqref{Psi^(1)-equation},
we also have a solution $\widetilde{\Psi}^{(1)}$ to~\eqref{Psi-tilde^(1)-equation}.
It is obvious that we can proceed to higher orders in $g$,
and we conclude that when we have a perturbative solution~$\Psi$ to~\eqref{Psi-equation},
we have a solution~$\widetilde{\Psi}$ to~\eqref{Psi-tilde-equation}.
This implies that no further conditions are imposed on $\Psi$
when we solve~\eqref{Psi-tilde-equation}.
General solutions to~\eqref{Psi-tilde-equation} can be written as
\begin{equation}
\widetilde{\Psi} = \widetilde{\Psi}_\ast +\Delta \widetilde{\Psi} \,,
\end{equation}
where $\widetilde{\Psi}_\ast$ and $\Delta \widetilde{\Psi}$ satisfy, respectively,
\begin{align}
Q_{B} \widetilde{\Psi}_\ast & = -\sum_{n=2}^{\infty}\frac{g^{n-1}}{n!}[\Psi ^{n}] \,,
\label{Psi-tilde-star} \\
Q_{B} \Delta \widetilde{\Psi} & = 0 \,.
\label{Psi-tilde-fluctuation}
\end{align}
The fluctuation $\Delta \widetilde{\Psi}$ around $\widetilde{\Psi}_\ast$
obeys the free equation of motion, and it describes the spurious free fields.
To summarize, we have shown that the theory
described by the equations of motion~\eqref{delta-Psi-equation} and~\eqref{Psi-tilde-equation}
is equivalent to the theory described by the equation of morion~\eqref{c}
up to spurious free fields described by the fluctuation of $\widetilde{\Psi}$.

Finally, let us discuss the invariance of the action
under nonlinearly extended gauge transformations.
In the approach by Sen for the Ramond sector,
the nonlinear terms in the gauge transformations
for $\Psi$ of picture number~$-1/2$ are written
in terms of the multi-string products with a factor of $X$
and the nonlinear terms in the gauge transformations
for $\widetilde{\Psi}$ of picture number~$-3/2$ are written
in terms of the multi-string products without $X$.
Based on the analogy with the approach by Sen,
we use the multi-string products with a factor of $B$
for nonlinear terms in the gauge transformations
for $\Psi$ of ghost number $2$
and the multi-string products without $B$
for nonlinear terms in the gauge transformations
for $\widetilde{\Psi}$ of ghost number $3$
in our current construction of closed string field theory.
Consider the gauge transformations given by
\begin{align}
\delta _{\Lambda}\Psi &=Q_{B}\Lambda +\sum_{n=1}^{\infty}\frac{g^{n}}{n!}B[\Psi ^{n},\Lambda ]
=Q_{B}\Lambda +gB[\Psi ,\Lambda ]+\frac{g^{2}}{2!}B[\Psi ,\Psi ,\Lambda ]+O(g^{3}),&\label{aa} \\
\delta _{\Lambda}\widetilde{\Psi} &=-\sum_{n=1}^{\infty}\frac{g^{n}}{n!}[\Psi ^{n},\Lambda ]
=-g[\Psi ,\Lambda ]-\frac{g^{2}}{2!}[\Psi ,\Psi ,\Lambda ]+O(g^{3}),& \label{m} \\
\delta _{\widetilde{\Lambda}}\Psi  &=0,&\label{bbb} \\
\delta _{\widetilde{\Lambda}}\widetilde{\Psi}& =Q_{B}\widetilde{\Lambda},&\label{cc}
\end{align}
where $\Lambda$ and $\widetilde{\Lambda}$ are gauge parameters.
The action \eqref{d} is invariant under these gauge transformations
if the multi-string products satisfy 
\begin{eqnarray}
\begin{split}
0=&Q_{B}[A_{1},\cdots ,A_{n}]-\sum_{i=1}^{n}(-1)^{(A_{1}+\cdots +A_{i-1})}[A_{1},\cdots ,Q_{B}A_{i},\cdots ,A_{n}]&\\
&-\sum_{\substack{\{ i_{l},j_{k}\}\\ l\geq 1,k\geq 2 }}\sigma ( i_{l},j_{k})[A_{i_{1}},\cdots ,A_{i_{l}},B[A_{j_{1}}\cdots ,A_{j_{k}}]].& \label{h}
\end{split}
\end{eqnarray}
As in~\eqref{g},
the second summation on the right-hand side runs over all different splittings of the set $\{ 1,\cdots , n\}$ into a first group $\{ i_{1},\cdots , i_{l}\}$ and a second group $\{ j_{1},\cdots , j_{k}\} $. The sign factor $\sigma ( i_{l},j_{k})$ is the sign picked up when we rearrange the sequence $\{ Q_{B},A_{1},\cdots ,A_{n}\} $ into the sequence $\{ A_{i_{1}},\cdots ,A_{i_{l}},Q_{B},A_{j_{1}},\cdots ,A_{j_{k}}\}$ taking into account the Grassmann property of the various objects.

The relations~\eqref{g}
for the multi-string products $[\![A_{1},\cdots ,A_{n} ]\!]$
can be derived from the relations~\eqref{h} for $[A_{1},\cdots ,A_{n} ]$
using $[\![ A_{1},A_{2},\cdots ,A_{n}]\!] = B \, [ A_{1},A_{2},\cdots ,A_{n}]$
in \eqref{identification-of-string-products}.
In fact, when we construct the multi-string products $[\![ A_{1},A_{2},\cdots ,A_{n}]\!]$
satisfying the relations~\eqref{g}
based on the decomposition of the moduli space of Riemann surfaces,
multi-string products $[A_{1},\cdots ,A_{n} ]$
satisfying~\eqref{h} naturally appear
at an intermediate step,
and the multi-string products $[\![ A_{1},A_{2},\cdots ,A_{n}]\!]$
of the form~\eqref{double-bracket-from-F}
are constructed from
the multi-string products $[A_{1},\cdots ,A_{n} ]$ of the form~\eqref{single-bracket-from-F}
satisfying~\eqref{h} as
$[\![ A_{1},A_{2},\cdots ,A_{n}]\!] = b_0^- \, [ A_{1},A_{2},\cdots ,A_{n}]$.

Let us see the invariance of the action up to $O(g^{3})$ under the gauge transformations explicitly. First of all, the variation $\delta _{\widetilde{\Lambda}} S$ vanishes because of the nilpotency of $Q_{B}$. Let us next consider the variation
with the parameter $\Lambda$:
\begin{eqnarray}
\begin{split}
\delta _{\Lambda} S&=\braket{\delta _{\Lambda}\widetilde{\Psi},Q_{B}B\widetilde{\Psi}}+\braket{\delta _{\Lambda}\widetilde{\Psi},Q_{B}\Psi}+\braket{\delta _{\Lambda}\Psi ,Q_{B}\widetilde{\Psi}}&\\
&\ \ \ \ \ \ \ \ \ \ \ \ \ \ \ \ \ \ \ \ \ \  +\frac{g}{2!}\braket{\delta _{\Lambda} \Psi ,[\Psi ,\Psi ]}+\frac{g^{2}}{3!}\braket{\delta _{\Lambda} \Psi ,[\Psi ,\Psi ,\Psi ]}+O(g^{3}) \,. &\label{ee}
\end{split}
\end{eqnarray}
As the nonlinear terms of $\delta _{\Lambda}\Psi $ in~\eqref{aa}
are written in terms of the multi-string products $[ A_{1},A_{2},\cdots ,A_{n}]$
with a factor of $B$
and the nonlinear terms of $\delta _{\Lambda}\widetilde{\Psi} $ in~\eqref{m}
are written in terms of $[ A_{1},A_{2},\cdots ,A_{n}]$
without $B$,
we find
\begin{eqnarray}
\begin{split}
\delta _{\Lambda}\Psi +B\delta _{\Lambda}\widetilde{\Psi} =Q_{B}\Lambda \,. \label{ddd}
\end{split}
\end{eqnarray}
Using this relation and the nilpotency of $Q_{B}$,
we can show that
the first term and the third term on the right-hand side of~\eqref{ee} cancel:
\begin{eqnarray}
\begin{split}
\braket{\delta _{\Lambda}\widetilde{\Psi},Q_{B}B\widetilde{\Psi}}+\braket{\delta _{\Lambda}\Psi ,Q_{B}\widetilde{\Psi}}&=\braket{\delta _{\Lambda}\Psi +B\delta _{\Lambda}\widetilde{\Psi},Q_{B}\widetilde{\Psi}}&\\
&=\braket{Q_{B}\Lambda  ,Q_{B}\widetilde{\Psi}}& \\
&=0 \,. & \label{eee}
\end{split}
\end{eqnarray}
Let us expand the action $(\ref{d})$ in $g$ as
\begin{equation}
 S = S^{(0)}+g\, S^{(1)}+g^{2}\, S^{(2)}+O(g^{3}), \label{ff}
\end{equation}
where
\begin{equation}
S^{(0)}= \frac{1}{2}\braket{\widetilde{\Psi},Q_{B}B\widetilde{\Psi}}+\braket{\widetilde{\Psi},Q_{B}\Psi},\ \ S^{(1)}=\frac{1}{3!}\braket{\Psi ,[\Psi ,\Psi ]},\ \  S^{(2)}=\frac{1}{4!}\braket{ \Psi ,[\Psi ,\Psi ,\Psi ]}.
\end{equation}
We have seen from~\eqref{eee}
that the variation $\delta_\Lambda S^{(0)}$ is given by
\begin{equation}
  \delta _{\Lambda} S^{(0)} =\braket{\delta _{\Lambda}\widetilde{\Psi},Q_{B}\Psi} \,.
\end{equation}
Let us also expand
the gauge transformations~$(\ref{aa})$ and~$(\ref{m})$ in $g$ as
  \begin{align}
\delta _{\Lambda}\Psi &=\delta _{\Lambda}^{(0)}\Psi +g\, \delta _{\Lambda}^{(1)}\Psi +g^{2}\, \delta _{\Lambda}^{(2)}\Psi +O(g^{3}),&\\
\delta _{\Lambda}\widetilde{\Psi} &=g\, \delta _{\Lambda}^{(1)}\widetilde{\Psi} +g^{2}\, \delta _{\Lambda}^{(2)}\widetilde{\Psi} +O(g^{3}),&
 \end{align}
where 
 \begin{eqnarray}
  \begin{split}
&\delta _{\Lambda}^{(0)}\Psi =Q_{B}\Lambda ,\ \ \ \ \ \ \ \ \  \ \ \ \ \ \ \  \delta _{\Lambda}^{(1)}\Psi =B[ \Psi ,\Lambda ] ,\ \ \ \ \ \ \  \ \ \ \ \delta _{\Lambda}^{(2)}\Psi =\frac{1}{2!}B[ \Psi ,\Psi ,\Lambda ] .&\\
&\delta _{\Lambda}^{(1)}\widetilde{\Psi} =-[ \Psi ,\Lambda ] ,\ \ \ \ \ \ \  \ \ \ \ \delta _{\Lambda}^{(2)}\widetilde{\Psi} =-\frac{1}{2!}[ \Psi ,\Psi ,\Lambda ] .&
 \end{split}
\end{eqnarray}
The variation $\delta_\Lambda S$ at $O(g)$ is given by
 \begin{eqnarray}
  \begin{split}
\delta _{\Lambda}^{(1)} S^{(0)}+\delta _{\Lambda}^{(0)} S^{(1)}&=-\braket{[\Psi ,\Lambda ],Q_{B}\Psi}+\frac{1}{2!}\braket{ Q_{B}\Lambda ,[\Psi ,\Psi ]}&\\
&=\frac{1}{2!}\braket{ \Lambda ,Q_{B}[\Psi ,\Psi ]}-\braket{\Lambda ,[Q_{B}\Psi ,\Psi ]}.&\label{o}
\end{split}
\end{eqnarray}
The gauge invariance at this order follows from
\begin{eqnarray}
\begin{split}
0=Q_{B}[A_{1},A_{2}]-[Q_{B}A_{1},A_{2}]-(-1)^{A_{1}}[A_{1},Q_{B}A_{2}],
\end{split}
\end{eqnarray}
which is the relation $(\ref{h})$ for $n=2$. The variation at $O(g^{2})$ is
 \begin{eqnarray}
  \begin{split}
 &\delta _{\Lambda}^{(2)} S^{(0)}+\delta _{\Lambda}^{(1)} S^{(1)}+\delta _{\Lambda}^{(0)} S^{(2)}&\\
 =&\frac{1}{3!}\braket{Q_{B}\Lambda  ,[\Psi ,\Psi ,\Psi ]}-\frac{1}{2!}\braket{[\Psi ,\Psi ,\Lambda ],Q_{B}\Psi}+\frac{1}{2!}\braket{B[\Psi ,\Lambda ],[\Psi ,\Psi ]}& \\
=&\frac{1}{3!}\braket{\Lambda  ,Q_{B}[\Psi ,\Psi ,\Psi ]} -\frac{1}{2!}\braket{\Lambda ,[Q_{B}\Psi ,\Psi ,\Psi ]}-\frac{1}{2!}\braket{\Lambda ,[\Psi ,B[\Psi ,\Psi ]]}.&
\label{p}
\end{split}
\end{eqnarray}
The gauge invariance at this order follows from
\begin{eqnarray}
0=Q_{B}[A_{1},A_{2},A_{3}]-[Q_{B}A_{1},A_{2},A_{3}]-(-1)^{A_{1}}[A_{1},Q_{B}A_{2},A_{3}]-(-1)^{A_{1}+A_{2}}[A_{1},A_{2},Q_{B}A_{3}]  \nonumber \\
-(-1)^{A_{1}}[A_{1},B[A_{2},A_{3}]]-(-1)^{A_{2}(1+A_{1})}[A_{2},B[A_{1},A_{3}]]-(-1)^{A_{3}(1+A_{1}+A_{2})}[A_{3},B[A_{1},A_{2}]],\nonumber \\ 
\end{eqnarray}
which is the relation $(\ref{h})$ for $n=3$.

\section{Expansion in terms of component fields}\label{expansion}
\setcounter{equation}{0}
In this section we study
closed string field theory
without the level-matching condition
by expanding the string fields
in terms of component fields
and confirm the equivalence to closed string field theory
with the level-matching condition
up to extra free fields.

We consider closed strings in a flat spacetime of 26 dimensions.\footnote{
We use $\mu$, $\nu$, and $\rho$ to label spacetime directions:
$\mu, \nu, \rho = 0, 1, \ldots, 25$.
}
The Hilbert space of the closed string is constructed
in terms of the operators
$\alpha^\mu_n$, $\widetilde{\alpha}^\mu_n$, $c_n$, $\widetilde{c}_n$,
$b_n$, and $\widetilde{b}_n$ satisfying
\begin{align}
[ \, \alpha^\mu_m,\alpha^\nu_n \, ] = m \, \delta_{m,-n} \eta^{\mu \nu} \,, \qquad
[ \, \widetilde{\alpha}^{\, \mu}_m,\widetilde{\alpha}^{\, \nu}_n \, ] = m \, \delta_{m,-n} \eta^{\mu \nu} \\
\{ \, c_m, b_n \, \} = \delta_{m,-n} \,, \qquad
\{ \, \widetilde{c}_m, \widetilde{b}_n \, \} = \delta_{m,-n} \,,
\end{align}
and we define the state $\ket{0;k}$ carrying spacetime momentum $k^\mu$ by
\begin{align}
& \alpha^\mu_n \ket{0;k} = 0 \quad \text{for} \quad n \ge 1 \,, \qquad
\widetilde{\alpha}^{\, \mu}_n \ket{0;k} = 0  \quad \text{for} \quad n \ge 1 \,, \\
& c_n \ket{0;k} = 0 \quad \text{for} \quad n \ge 2 \,, \qquad
\widetilde{c}_n \ket{0;k} = 0  \quad \text{for} \quad n \ge 2 \,, \\
& b_n \ket{0;k} = 0 \quad \text{for} \quad n \ge -1 \,, \qquad
\widetilde{b}_n \ket{0;k} = 0  \quad \text{for} \quad n \ge -1 \,.
\end{align}
We denote the creation-annihilation normal ordering of $\mathcal{O}$
by ${}^{\circ}_{\circ}\ \mathcal{O} \ ^{\circ}_{\circ} \,$.\footnote{
The creation-annihilation normal ordering
for the Grassmann-odd operators $c_n$, $\widetilde{c}_n$,
$b_n$, and $\widetilde{b}_n$
is defined by
\begin{equation*}
\begin{split}
\ ^{\circ}_{\circ}\ c_n b_{-n} \ ^{\circ}_{\circ}\ = \biggl\{
\begin{array}{lcl}
c_n b_{-n} & \text{for} & n \le 0  \,, \\
{}-b_{-n} c_n & \text{for} & n > 0 \,,
\end{array}
\\
\ ^{\circ}_{\circ}\ \widetilde{c}_n \widetilde{b}_{-n} \ ^{\circ}_{\circ}\ = \biggl\{
\begin{array}{lcl}
\widetilde{c}_n \widetilde{b}_{-n} & \text{for} & n \le 0  \,, \\
{}-\widetilde{b}_{-n} \widetilde{c}_n & \text{for} & n > 0 \,.
\end{array}
\end{split}
\end{equation*}
}
The BRST operator $Q_{B}$ is given by
\begin{eqnarray}
\begin{split}
Q_{B}=&\sum_{n=-\infty}^{\infty}(c_{n}L_{-n}^{(m)}+\widetilde{c}_{n}\widetilde{L}_{-n}^{(m)}) &\\
&+\frac{1}{2}\sum_{n=-\infty}^{\infty}\sum_{m=-\infty}^{\infty}(m-n)  \ ^{\circ}_{\circ}\ (c_{m}c_{n}b_{-m-n}+\widetilde{c}_{m}\widetilde{c}_{n}\widetilde{b}_{-m-n})\ ^{\circ}_{\circ}-(c_{0}+\widetilde{c}_{0}),&
\end{split}
\end{eqnarray}
where $L_{n}^{(m)}$ and $\widetilde{L}_{n}^{(m)}$ are
\begin{eqnarray}
\begin{split}
L_{n}^{(m)}&=\frac{1}{2}\sum_{m=-\infty}^{\infty}  \ ^{\circ}_{\circ}\ \alpha ^{\mu}_{m}\alpha ^{\nu}_{n-m}\ ^{\circ}_{\circ}\ \eta _{\mu \nu},&\\
\widetilde{L}_{n}^{(m)}&=\frac{1}{2}\sum_{m=-\infty}^{\infty}  \ ^{\circ}_{\circ}\ \widetilde{\alpha }^{\mu}_{m}\widetilde{\alpha }^{\nu}_{n-m}\ ^{\circ}_{\circ}\ \eta _{\mu \nu}.&
\end{split}
\end{eqnarray}
We normalize the BPZ inner product of $c_{-1}\widetilde{c}_{-1}c_{0}^{-}c_{0}^{+}c_{1}\widetilde{c}_{1}\ket{0;k}$
and $\ket{0;k'}$ as
\begin{eqnarray}
\bra{0;k^{\prime}}c_{-1}\widetilde{c}_{-1}c_{0}^{-}c_{0}^{+}c_{1}\widetilde{c}_{1}\ket{0;k}=2(2\pi )^{26}\delta ^{26}(k+k^{\prime}),
\end{eqnarray}
where $c_{0}^{-}$ and $c_{0}^{+}$ are defined by
\begin{eqnarray}
 \begin{split}
c_{0}^{-} &= \frac{1}{2}\, (c_{0}-\widetilde{c}_{0}),&\\
c_{0}^{+} &= \frac{1}{2}\, (c_{0}+\widetilde{c}_{0}).&
 \end{split}
\end{eqnarray}

In expanding the string fields,
we use a basis which consists of
eigenstates of $L_0$ and $\widetilde{L}_0$.
When the eigenvalues of $L_0$ and $\widetilde{L}_0$
for a state in the basis carrying spacetime momentum $k_\mu$
are $-1+\ell +\alpha' k^2 / 4$ and $-1+\tilde{\ell} +\alpha' k^2 / 4$,
respectively,
we say that the level of the state is $( \, \ell,\, \tilde{\ell} \, )$.
The string fields $\Psi$ and $\widetilde{\Psi}$
and the gauge parameters $\Lambda$ and $\widetilde{\Lambda}$
can be expanded with respect to the level as
\begin{align}
\Psi &= \sum_{\ell \, , \, \widetilde{\ell} \, = \, 0}^\infty \Psi _{( \, \ell \, , \, \tilde{\ell} \, )}\, , \qquad
\widetilde{\Psi} = \sum_{\ell \, , \, \widetilde{\ell} \, = \, 0}^\infty \widetilde{\Psi }_{( \, \ell \, , \, \tilde{\ell} \, )}\, , \\
\Lambda &= \sum_{\ell \, , \, \widetilde{\ell} \, = \, 0}^\infty \Lambda _{( \, \ell \, , \, \tilde{\ell} \, )}\, , \qquad
\widetilde{\Lambda} = \sum_{\ell \, , \, \widetilde{\ell} \, = \, 0}^\infty \widetilde{\Lambda }_{( \, \ell \, , \, \tilde{\ell} \, )}\, ,
\end{align}
where $\Psi _{( \, \ell \, , \, \widetilde{\ell} \, )} $,
$\widetilde{\Psi }_{( \, \ell \, , \, \tilde{\ell} \, )}$,
$\Lambda _{( \, \ell \, , \, \widetilde{\ell} \, )} $, and $\widetilde{\Lambda }_{( \, \ell \, , \, \tilde{\ell} \, )}$
consist of states of the level $( \, \ell,\, \tilde{\ell} \, )$.
In subsection~\ref{(0,0)} we study the sector of the level $(0,0)$,
where the component fields are tachyonic.
In subsection~\ref{(0,1)} we consider the sector of the level $(0,1)$.
All component fields in this sector violate the level-matching condition. 
In subsection~\ref{(1,1)} we consider the sector of the level $(1,1)$.
This sector contains massless fields including the graviton.

\subsection{Tachyonic fields}
\label{(0,0)}
The string field~$\Psi_{(0,0)}$ carrying ghost number $2$
is expanded as
\begin{eqnarray}
\Psi _{(0,0)}=\int \frac{d^{26}k}{(2\pi )^{26}}\ T(k) \, c_{1}\widetilde{c}_{1}\ket{0;k} \,,
\end{eqnarray}
where $T(k)$ is the closed string tachyon field.
There are no states of the level $(0,0)$
carrying ghost number $1$ so that the gauge parameter $\Lambda _{(0,0)}$ vanishes:
\begin{equation}
\Lambda _{(0,0)} = 0 \,.
\end{equation}
Therefore, there are no gauge symmetries associated
with the closed string tachyon $T(k)$ in the free theory.

The string field~$\widetilde{\Psi}_{(0,0)}$ carrying ghost number $3$
is expanded as
\begin{eqnarray}
\widetilde{\Psi}_{(0,0)} = \int \frac{d^{26}k}{(2\pi )^{26}}\, \big[\,  \widetilde{S}(k)c_{0}^{+} c_{1}\widetilde{c}_{1}\ket{0;k}+\widetilde{T}(k)c_{0}^{-} c_{1}\widetilde{c}_{1}\ket{0;k}\big] \,,
\end{eqnarray}
where $\widetilde{S}(k)$ and $\widetilde{T}(k)$ are two component fields at this level.
The gauge parameter $\widetilde{\Lambda }_{(0,0)}$ is expanded as
\begin{eqnarray}
\widetilde{\Lambda }_{(0,0)}=\int \frac{d^{26}k}{(2\pi )^{26}}\ \widetilde{\lambda} (k) \, c_{1}\widetilde{c}_{1}\ket{0;k} \,,
\end{eqnarray}
where $\widetilde{\lambda} (k)$ is the only one component field at this level.
The gauge transformation
\begin{equation}
\delta_{\widetilde{\Lambda}} \widetilde{\Psi}_{(0,0)}
= Q_B \widetilde{\Lambda}_{(0,0)}
\end{equation}
is expanded in terms of component fields as
\begin{eqnarray}
\delta _{\widetilde{\Lambda}}\widetilde{S}(k)=2\bigg( \frac{\alpha ^{\prime}k^{2}}{4}-1\bigg)\widetilde{\lambda} (k). \label{3000}
\end{eqnarray}

Let us next expand the kinetic terms.
Since
\begin{eqnarray}
\begin{split}
\braket{\widetilde{\Psi}_{(0,0)},Q_{B}B\widetilde{\Psi}_{(0,0)} }&=-\int \frac{d^{26}k}{(2\pi )^{26}}\, \widetilde{T}(-k)\bigg( \frac{\alpha ^{\prime}k^{2}}{4}-1\bigg) \widetilde{T}(k)
\end{split}
\end{eqnarray}
and
\begin{eqnarray}
\begin{split}
\braket{\widetilde{\Psi}_{(0,0)},Q_{B}\Psi_{(0,0)}}&=-\int \frac{d^{26}k}{(2\pi )^{26}}\, \widetilde{T}(-k)\bigg( \frac{\alpha ^{\prime}k^{2}}{4}-1\bigg) T(k) \,,
\end{split}
\end{eqnarray}
the kinetic terms of the level $(0,0)$ are given by
\begin{eqnarray}
-\int \frac{d^{26}k}{(2\pi )^{26}}\bigg[ \, \frac{1}{2}\widetilde{T}(-k)\bigg( \frac{\alpha ^{\prime}k^{2}}{4}-1\bigg) \, \widetilde{T}(k)+\, \widetilde{T}(-k)\bigg( \frac{\alpha ^{\prime}k^{2}}{4}-1\bigg) T(k)\bigg] . 
\end{eqnarray}
Note that the field $\widetilde{S}(k)$ does not appear in the kinetic terms.
Therefore, the kinetic terms are trivially invariant
under the gauge transformation~\eqref{3000}.\footnote{
In the context of the BRST cohomology on states of ghost number $3$,
the state $c_{0}^{-} c_{1}\widetilde{c}_{1}\ket{0;k}$ 
corresponding to $\widetilde{T}(k)$
is BRST closed when $\alpha' k^2 /4 -1 = 0$
and it is not BRST exact.
On the other hand,
the state $c_{0}^{+} c_{1}\widetilde{c}_{1}\ket{0;k}$ 
corresponding to $\widetilde{S}(k)$
is BRST closed for any $k_\mu$,
but it is BRST exact when $\alpha' k^2 /4 -1 \ne 0$.
This way we have two copies of on-shell tachyon states
represented by $c_{0}^{-} c_{1}\widetilde{c}_{1}\ket{0;k}$
with $\alpha' k^2 /4 -1 = 0$
and $c_{0}^{+} c_{1}\widetilde{c}_{1}\ket{0;k}$
with $\alpha' k^2 /4 -1 = 0$.
In the context of string field theory,
the component field $\widetilde{S}(k)$ does not appear
in the action
and we have only one tachyon from $\widetilde{\Psi}$ of ghost number $3$.
}

The equations of motion
\begin{align}
Q_{B}B\widetilde{\Psi}+Q_{B}\Psi & = 0 \,, \\
Q_{B}\widetilde{\Psi}+\sum_{n=2}^{\infty}\frac{g^{n-1}}{n!}[ \, \Psi ^{n} \, ] & = 0
\label{Psi-tilde-equation-2}
\end{align}
are expanded in terms of component fields at this level as
\begin{align}
\bigg( \frac{\alpha ^{\prime}k^{2}}{4}-1\bigg) \, \widetilde{T}(k)+\bigg( \frac{\alpha ^{\prime}k^{2}}{4}-1\bigg) T(k)&=0,& \\
\bigg( \frac{\alpha ^{\prime}k^{2}}{4}-1\bigg) \, \widetilde{T}(k)&=\mathcal{J}_{T}(k),&
\label{Psi-tilde-(0,0)}
\end{align}
where the source term $\mathcal{J}_{T}(k)$ is from the interaction terms
which depend only on the string field $\Psi$.
After eliminating $\widetilde{\Psi}$,
the equation for $\Psi$~\eqref{Psi-equation},
\begin{equation}
Q_{B}\Psi+\sum_{n=2}^{\infty}\frac{g^{n-1}}{n!} B \, [ \, \Psi ^{n} \, ] = 0 \,,
\label{Psi-equation-2}
\end{equation}
is expanded as
\begin{equation}
\bigg( \frac{\alpha ^{\prime}k^{2}}{4}-1\bigg) T(k) =-\mathcal{J}_{T}(k) \,.
\label{Psi-(0,0)}
\end{equation}
This coincides with the equation of motion for the closed string tachyon
in closed string field theory with the level-matching condition.
Once we have a solution to the equation for $\Psi$~\eqref{Psi-equation-2},
the equation~\eqref{Psi-tilde-equation-2} determines $\widetilde{\Psi}$,
which is expanded at this level as~\eqref{Psi-tilde-(0,0)}.
Following the general discussion in subsection~\ref{section-2.2},
the equation~\eqref{Psi-tilde-(0,0)} must have a solution for $\widetilde{T}(k)$
when we have a solution for $T(k)$ to the equation~\eqref{Psi-(0,0)}.
In order for~\eqref{Psi-(0,0)} to have a solution for $T(k)$,
the source term $\mathcal{J}_{T}(k)$ must vanish at $k^{2}=4/\alpha ^{\prime}$.
In this case the equation~\eqref{Psi-tilde-(0,0)} can be solved for $\widetilde{T}(k)$,
in accord with the general discussion.
When we denote the solution by $\widetilde{T}_\ast(k)$
and we expand $\widetilde{T}(k)$ as
\begin{equation}
\widetilde{T}(k) = \widetilde{T}_\ast(k) +\Delta \widetilde{T}(k) \,,
\end{equation}
the fluctuation $\Delta \widetilde{T}(k)$ obeys the equation of motion
\begin{align}
\bigg( \frac{\alpha ^{\prime}k^{2}}{4}-1\bigg) \, \Delta \widetilde{T}(k)=0 \,,
\label{free-T-tilde^(-)}
\end{align}
which is the extra free field at this level.

To summarize,
we find the closed string tachyon $T(k)$
and the extra field $\Delta \widetilde{T}(k)$ at the level $(0,0)$.
We confirm that $T(k)$ obeys the equation of motion
for the closed string tachyon~\eqref{Psi-(0,0)} 
and that $\Delta \widetilde{T}(k)$ is a free field
satisfying the equation of motion~\eqref{free-T-tilde^(-)}.

\subsection{Component fields violating the level-matching condition}
\label{(0,1)}
Let us next consider the level~$(0,1)$.
All component fields at this level violate the level-matching condition
so that this sector is absent
in closed string field theory with the level-matching condition.
The analysis at the level~$(1,0)$ is completely parallel
to the analysis in this subsection.

\subsubsection{The expansions in terms of component fields}

The string field $\Psi_{(0,1)}$ carrying ghost number $2$ is expanded as
\begin{equation}
\Psi _{(0,1)}=\int \frac{d^{26}k}{(2\pi )^{26}}\ \big[ \, K(k)\, c_{0}c_{1}\ket{0;k}+L(k)\, c_{1}\widetilde{c}_{0}\ket{0;k}+U_{\mu}(k)\, \widetilde{\alpha} ^{\mu}_{-1}c_{1}\widetilde{c}_{1}\ket{0;k}\big] \, ,
\end{equation}
where $K(k)$, $L(k)$, and $U_{\mu}(k)$ are three component fields from $\Psi$ at this level.
The gauge parameter $\Lambda_{(0,1)} $ is expanded as
\begin{equation}
\Lambda _{(0,1)}=\int \frac{d^{26}k}{(2\pi )^{26}}\ \kappa (k)\, c_{1}\ket{0;k}\, ,
\end{equation}
where $\kappa (k)$ is a component field from $\Lambda$ at this level.
The gauge transformation
\begin{equation}
\delta _{\Lambda}\Psi_{(0,1)} =Q_{B}\Lambda _{(0,1)}
\end{equation}
is expanded in terms of component fields as
\begin{equation}
\delta _{\Lambda}K(k)=\bigg( \frac{\alpha ^{\prime}k^{2}}{4}-1\bigg)  \kappa (k) \, , \quad 
\delta _{\Lambda}L(k)=-\frac{\alpha ^{\prime}k^{2}}{4}\kappa (k)\, , \quad
\delta _{\Lambda}U_{\mu}(k)=-\sqrt{\frac{\alpha ^{\prime}}{2}}k_{\mu}\kappa  (k)\, .\label{3}
\end{equation}

The string field $\widetilde{\Psi}_{(0,1)}$ carrying ghost number $3$ is expanded as
\begin{equation}
\begin{split}
\widetilde{\Psi} _{(0,1)}&=\int \frac{d^{26}k}{(2\pi )^{26}}\ \big[ \, \widetilde{M}(k)\, c_{0}c_{1}\widetilde{c}_{0}\ket{0;k}+ \widetilde{N}(k)\, c_{1}\widetilde{c}_{-1}\widetilde{c}_{1}\ket{0;k}\\
&\ \ \ \ \ \ \ \ \ \ \ \ \ \ \ \ \ \ \ \ \ \ \ \ \ \ \ \ \ + \widetilde{V}_{\mu}(k)\, \widetilde{\alpha} ^{\mu}_{-1}c_{0}c_{1}\widetilde{c}_{1}\ket{0;k}+ \widetilde{W}_{\mu}(k)\, \widetilde{\alpha} ^{\mu}_{-1}c_{1}\widetilde{c}_{0}\widetilde{c}_{1}\ket{0;k} \big] \, ,
\end{split}
\end{equation}
where $\widetilde{M}(k)$, $\widetilde{N}(k)$, $\widetilde{V}_{\mu}(k)$, and $\widetilde{W}_{\mu}(k)$ are four spacetime fields from $\widetilde{\Psi}$ at this level.
The gauge parameter $\widetilde{\Lambda }_{(0,1)}$ is expanded as
\begin{equation}
\widetilde{\Lambda }_{(0,1)}=\int \frac{d^{26}k}{(2\pi )^{26}}\ \big[ \, \widetilde{\beta}(k)\, c_{0}c_{1}\ket{0;k}+\widetilde{\gamma}(k)\, c_{1}\widetilde{c}_{0}\ket{0;k}+\widetilde{\theta} _{\mu}(k)\, \widetilde{\alpha} ^{\mu}_{-1}c_{1}\widetilde{c}_{1}\ket{0;k}\big] \, .
\end{equation}
where $\widetilde{\beta}(k)$, $\widetilde{\gamma}(k)$, and $\widetilde{\theta} _{\mu}(k)$ are three component fields at this level.
The gauge transformation
\begin{equation}
\delta _{\widetilde{\Lambda}}\widetilde{\Psi} _{(0,1)}=Q_{B}\widetilde{\Lambda }_{(0,1)}\, ,
\end{equation}
is expanded in terms of component fields as
\begin{equation}
\begin{split}
\delta _{\Lambda}\widetilde{M}(k)&=\frac{\alpha ^{\prime}k^{2}}{4}\widetilde{\beta}(k)+\bigg( \frac{\alpha ^{\prime}k^{2}}{4}-1\bigg) \widetilde{\gamma}(k)\, , \\  
\delta _{\Lambda}\widetilde{N}(k)&=2\widetilde{\gamma}(k)-\sqrt{\frac{\alpha ^{\prime}}{2}}k^{\mu}\widetilde{\theta} _{\mu}(k)\, , \\ 
\delta _{\Lambda}\widetilde{V}_{\mu}(k)&=\sqrt{\frac{\alpha ^{\prime}}{2}}k_{\mu}\widetilde{\beta}(k)+\bigg( \frac{\alpha ^{\prime}k^{2}}{4}-1\bigg) \widetilde{\theta} _{\mu}(k)\, , \\
\delta _{\Lambda}\widetilde{W}_{\mu}(k)&=\sqrt{\frac{\alpha ^{\prime}}{2}}k_{\mu}\widetilde{\gamma}(k)-\frac{\alpha ^{\prime}k^{2}}{4}\widetilde{\theta} _{\mu}(k)\, .\label{9}
\end{split}
\end{equation}
Note that gauge transformations parameterized
by $\widetilde{\beta} (k)$, $\widetilde{\gamma} (k)$, and $\widetilde{\theta}_{\mu}(k)$
are not independent
and the gauge transformation of the form
\begin{equation}
\widetilde{\Lambda }_{(0,1)} = Q_B \widetilde{\Omega }_{(0,1)}
\end{equation}
with
\begin{equation}
\widetilde{\Omega} _{(0,1)}=\int \frac{d^{26}k}{(2\pi )^{26}}\ \widetilde{\sigma} (k)\, c_{1}\ket{0;k} \,,
\end{equation}
where $\widetilde{\sigma} (k)$ is a component field,
does not change $\widetilde{\Psi}_{(0,1)}$ because $Q_B^2 = 0$.
In terms of component fields, this can be expressed as
\begin{equation}
\widetilde{\beta} (k)=\bigg( \frac{\alpha ^{\prime}k^{2}}{4}-1\bigg)  \widetilde{\sigma} (k) \, , \quad 
\widetilde{\gamma} (k)=-\frac{\alpha ^{\prime}k^{2}}{4}\widetilde{\sigma} (k)\, , \quad
\widetilde{\theta}_{\mu}(k)=-\sqrt{\frac{\alpha ^{\prime}}{2}}k_{\mu}\widetilde{\sigma}  (k)\, .\label{degree-of-freedoms-of-gauge-parameters}
\end{equation}

Let us next expand the kinetic terms. The kinetic terms at this level simplify as
\begin{equation}
\frac{1}{2}\braket{\widetilde{\Psi}_{(0,1)},Q_{B}B\widetilde{\Psi}_{(0,1)}}
+\braket{\widetilde{\Psi}_{(0,1)},Q_{B}\Psi_{(0,1)}}
= \braket{\widetilde{\Psi}_{(0,1)},Q_{B}\Psi _{(0,1)}} \label{hhh}
\end{equation}
because $\widetilde{\Psi}_{(0,1)}$ does not satisfy the level-matching condition
and is annihilated by $B$:
\begin{equation}
B\widetilde{\Psi}_{(0,1)}=0\, .\label{violating the level-matching condition}
\end{equation}
The remaining term $\braket{\widetilde{\Psi}_{(0,1)},Q_{B}\Psi _{(0,1)}}$
is expanded as
\begin{equation}
\begin{split}
&\int \frac{d^{26}k}{(2\pi )^{26}}\ \bigg\{ \widetilde{M}(-k)\bigg[ 2L(k)-\sqrt{\frac{\alpha ^{\prime}}{2}}k^{\mu}U_{\mu}(k)\bigg] \\
&\ \ \ \ \ \ \ \ \ \ \ \ \ \ \ \   -\widetilde{N}(-k)\bigg[ \frac{\alpha ^{\prime}k^{2}}{4}K(k)+\bigg( \frac{\alpha ^{\prime}k^{2}}{4}-1\bigg)L(k)\bigg] \\
&\ \ \ \ \ \ \ \ \ \ \ \ \ \ \ \  -\widetilde{W}^{\mu}(-k)\bigg[ \sqrt{\frac{\alpha ^{\prime}}{2}}k_{\mu}K(k)+\bigg( \frac{\alpha ^{\prime}k^{2}}{4}-1\bigg) U_{\mu}(k) \bigg] \\
&\ \ \ \ \ \ \ \ \ \ \ \ \ \ \ \  +\widetilde{V}^{\mu}(-k)\bigg[ \sqrt{\frac{\alpha ^{\prime}}{2}}k_{\mu}L(k)-\frac{\alpha ^{\prime}k^{2}}{4}U_{\mu}(k)\bigg] \bigg\} \, . \\
\end{split}
\end{equation}

Since the interaction terms do not contain $\widetilde{\Psi}_{(0,1)}$,
the equation of motion when we vary the action with respect to $\widetilde{\Psi}_{(0,1)}$
is given by
\begin{equation}
Q_{B}\Psi_{(0,1)}  = 0 \,. \label{equation_Psi_(0,1)}
\end{equation}
This is written in terms of component fields as follows:
\begin{equation}
\begin{split}
2L(k)-\sqrt{\frac{\alpha ^{\prime}}{2}}k^{\mu}U_{\mu}(k)&=0 \,,\\
 \frac{\alpha ^{\prime}k^{2}}{4}K(k)+\bigg( \frac{\alpha ^{\prime}k^{2}}{4}-1\bigg)L(k)&=0 \,,\\
 \sqrt{\frac{\alpha ^{\prime}}{2}}k_{\mu}K(k)+\bigg( \frac{\alpha ^{\prime}k^{2}}{4}-1\bigg) U_{\mu}(k)&=0 \,,\\
\sqrt{\frac{\alpha ^{\prime}}{2}}k_{\mu}L(k)-\frac{\alpha ^{\prime}k^{2}}{4}U_{\mu}(k)&=0 \,.
\end{split}
\label{1}
\end{equation}
The equation of motion when we vary the action
with respect to $\Psi$,
\begin{equation}
Q_{B}\widetilde{\Psi}+\sum_{n=2}^{\infty}\frac{g^{n-1}}{n!}[ \, \Psi ^{n} \, ]  = 0 \,,\label{equation_tilde_Psi_(0,1)}
\end{equation}
is expanded in terms of component fields at this level as
\begin{equation}
\begin{split}
 \frac{\alpha ^{\prime}k^{2}}{4}\widetilde{N}(k)- \sqrt{\frac{\alpha ^{\prime}}{2}}k_{\mu}\widetilde{W}^{\mu}(k)&=\mathcal{J}_{K}(k)\,,\\
 \bigg( \frac{\alpha ^{\prime}k^{2}}{4}-1\bigg) \widetilde{N}(k)-2\widetilde{M}(k)+ \sqrt{\frac{\alpha ^{\prime}}{2}}k_{\mu}\widetilde{V}^{\mu}(k)&=\mathcal{J}_{L}(k )\,, \\
\bigg( \frac{\alpha ^{\prime}k^{2}}{4}-1\bigg)\widetilde{W}^{\mu}(k)- \sqrt{\frac{\alpha ^{\prime}}{2}}k^{\mu}\widetilde{M}(k)+\frac{\alpha ^{\prime}k^{2}}{4}\widetilde{V}^{\mu}(k)&=\mathcal{J}^{\mu}_{U}(k)\,,
\end{split}
\label{6}
\end{equation}
where the source terms $\mathcal{J}_{K}(k)$, $\mathcal{J}_{L}(k)$ and $\mathcal{J}^{\mu}_{U}(k)$ are from the interaction terms.

\subsubsection{Solutions to the equations of motion}
\label{subsection-3.2.2}

Let us first solve the equations for motion~\eqref{1} for $K(k)$, $L(k)$, and $U_\mu (k)$.
Because of the invariance under the gauge transformation~\eqref{3}
we can choose the condition
\begin{equation}
\begin{split}
K(k) = 0 \quad \text{for} \quad k^2 & = 0 \,, \\
L(k) = 0 \quad \text{for} \quad k^2 & \ne 0
\end{split}
\end{equation}
to fix a gauge.
When $K(k) = 0$, the equations of motion~\eqref{1} simplify as follows:
\begin{equation}
\begin{split}
2L(k)-\sqrt{\frac{\alpha ^{\prime}}{2}}k^{\mu}U_{\mu}(k)&=0 \,,\\
\bigg( \frac{\alpha ^{\prime}k^{2}}{4}-1\bigg)L(k)&=0 \,,\\
\bigg( \frac{\alpha ^{\prime}k^{2}}{4}-1\bigg) U_{\mu}(k)&=0 \,,\\
\sqrt{\frac{\alpha ^{\prime}}{2}}k_{\mu}L(k)-\frac{\alpha ^{\prime}k^{2}}{4}U_{\mu}(k)&=0 \,.
\end{split}
\end{equation}
We choose the condition $K(k)=0$ when $k^2 = 0$
and in this case $\alpha^\prime k^2 /4 -1$ is nonvanishing
so that we find
\begin{equation}
L(k) = 0 \,, \qquad U_\mu (k) = 0 \,.
\end{equation}
When $L(k) = 0$, the equations of motion~\eqref{1} simplify as follows:
\begin{equation}
\begin{split}
-\sqrt{\frac{\alpha ^{\prime}}{2}}k^{\mu}U_{\mu}(k)&=0 \,,\\
 \frac{\alpha ^{\prime}k^{2}}{4}K(k)&=0 \,,\\
 \sqrt{\frac{\alpha ^{\prime}}{2}}k_{\mu}K(k)+\bigg( \frac{\alpha ^{\prime}k^{2}}{4}-1\bigg) U_{\mu}(k)&=0 \,,\\
-\frac{\alpha ^{\prime}k^{2}}{4}U_{\mu}(k)&=0 \,.
\end{split}
\end{equation}
We choose the condition $L(k)=0$ when $k^2 \ne 0$
and in this case we find
\begin{equation}
K(k) = 0 \,, \qquad U_\mu (k) = 0 \,.
\end{equation}
We thus conclude that all the solutions to the equations of motion~\eqref{1}
are equivalent to
\begin{equation}
K(k) = 0 \,, \qquad L(k) = 0 \,, \qquad U_\mu (k) = 0
\end{equation}
under the gauge transformation~\eqref{3}.

Let us next consider the equations~\eqref{6}.
The source terms $\mathcal{J}_{K}(k)$, $\mathcal{J}_{L}(k)$ and $\mathcal{J}^{\mu}_{U}(k)$
do not depend on $\widetilde{\Psi}$,
but they depend on component fields of $\Psi$ at all levels.
Once we have a solution to
\begin{equation}
Q_{B}\Psi+\sum_{n=2}^{\infty}\frac{g^{n-1}}{n!} B \, [ \, \Psi ^{n} \, ] = 0 \,,
\label{Psi-equation-3}
\end{equation}
the equations~\eqref{6} can be regarded as the equations
for $\widetilde{M}(k)$, $\widetilde{N}(k)$, $\widetilde{V}^{\mu}(k)$,
and $\widetilde{W}^{\mu}(k)$.
We will show in subsection~\ref{subsection-3.2.3} that
the equations~\eqref{6} can be solved
for the source terms from $\Psi$ satisfying~\eqref{Psi-equation-3},
and we denote the solutions by $\widetilde{M}_{\ast}(k)$, $\widetilde{N}_{\ast}(k)$, $\widetilde{V}^{\mu}_{\ast}(k)$ and $\widetilde{W}^{\mu}_{\ast}(k)$.
Here we consider the equations for fluctuations around the solutions,
which correspond to~\eqref{Psi-tilde-fluctuation}.
It is simpler to discuss the equation~\eqref{Psi-tilde-fluctuation}
than to solve the equation~\eqref{Psi-tilde-star} with source terms.

We expand $\widetilde{M}(k)$, $\widetilde{N}(k)$, $\widetilde{V}^{\mu}(k)$ and $\widetilde{W}^{\mu}(k)$ as
\begin{equation}
\begin{split}
\widetilde{M}(k)&=\widetilde{M}_{\ast}(k)+\Delta \widetilde{M}(k)\, ,\\
\widetilde{N}(k)&=\widetilde{N}_{\ast}(k)+\Delta \widetilde{N}(k)\, , \\
\widetilde{V}^{\mu}(k)&=\widetilde{V}^{\mu}_{\ast}(k)+\Delta \widetilde{V}^{\mu}(k)\, , \\
\widetilde{W}^{\mu}(k)&=\widetilde{W}^{\mu}_{\ast}(k)+\Delta \widetilde{W}^{\mu}(k)\, ,
\end{split}
\end{equation}
and the equations for the fluctuations $\Delta \widetilde{M}(k)$, $\Delta \widetilde{N}(k)$, $\Delta \widetilde{V}^{\mu}(k)$ and $\Delta \widetilde{W}^{\mu}(k)$ are given by
\begin{eqnarray}
\begin{split}
 \frac{\alpha ^{\prime}k^{2}}{4}\Delta \widetilde{N}(k)- \sqrt{\frac{\alpha ^{\prime}}{2}}k_{\mu}\Delta \widetilde{W}^{\mu}(k)&=0,&\\
 \bigg( \frac{\alpha ^{\prime}k^{2}}{4}-1\bigg) \Delta \widetilde{N}(k)-2\Delta \widetilde{M}(k)+ \sqrt{\frac{\alpha ^{\prime}}{2}}k_{\mu}\Delta \widetilde{V}^{\mu}(k)&=0, &\\
\bigg( \frac{\alpha ^{\prime}k^{2}}{4}-1\bigg)\Delta \widetilde{W}^{\mu}(k)- \sqrt{\frac{\alpha ^{\prime}}{2}}k^{\mu}\Delta \widetilde{M}(k)+\frac{\alpha ^{\prime}k^{2}}{4}\Delta \widetilde{V}^{\mu}(k)&=0.&\label{8}
 \end{split}
\end{eqnarray}
To fix a gauge for the gauge transformations
\begin{equation}
\begin{split}
\delta _{\Lambda} \Delta \widetilde{M}(k)&=\frac{\alpha ^{\prime}k^{2}}{4}\widetilde{\beta}(k)+\bigg( \frac{\alpha ^{\prime}k^{2}}{4}-1\bigg) \widetilde{\gamma}(k)\, , \\  
\delta _{\Lambda} \Delta \widetilde{N}(k)&=2\widetilde{\gamma}(k)-\sqrt{\frac{\alpha ^{\prime}}{2}}k^{\mu}\widetilde{\theta} _{\mu}(k)\, , \\ 
\delta _{\Lambda} \Delta \widetilde{V}_{\mu}(k)&=\sqrt{\frac{\alpha ^{\prime}}{2}}k_{\mu}\widetilde{\beta}(k)+\bigg( \frac{\alpha ^{\prime}k^{2}}{4}-1\bigg) \widetilde{\theta} _{\mu}(k)\, , \\
\delta _{\Lambda} \Delta \widetilde{W}_{\mu}(k)&=\sqrt{\frac{\alpha ^{\prime}}{2}}k_{\mu}\widetilde{\gamma}(k)-\frac{\alpha ^{\prime}k^{2}}{4}\widetilde{\theta} _{\mu}(k)\,,\end{split}
\label{gauge-transformations-for-fluctuations}
\end{equation}
we choose the conditions
\begin{equation}
\Delta\widetilde{N}(k)=0 \,, \quad
\Delta \widetilde{V}^{\mu}(k)=0 \quad \mathrm{for} \quad k^{2}=0
\end{equation}
and
\begin{equation}
\Delta\widetilde{M}(k)=0 \,, \quad
\Delta \widetilde{W}^{\mu}(k)=0\quad \mathrm{for} \quad k^{2}\ne 0\, .
\end{equation}
When $k^2 = 0$, the conditions
$\Delta\widetilde{N}(k)=0$
and
$\Delta \widetilde{V}^{\mu}(k)=0$
can be satisfied by the gauge transformations
with the parameters $\widetilde{\gamma}(k)$ and $\widetilde{\theta} _{\mu}(k)$.
When $k^2 \ne 0$, the conditions
$\Delta\widetilde{M}(k)=0$
and
$\Delta \widetilde{W}^{\mu}(k)=0$
can be satisfied by the gauge transformations
with the parameters $\widetilde{\beta}(k)$ and $\widetilde{\theta} _{\mu}(k)$.

Under the conditions
$\Delta\widetilde{V}(k)=0$ and $\Delta \widetilde{W}^{\mu}(k)=0$
the equations~\eqref{8} simplify as follows:
\begin{eqnarray}
\begin{split}
\sqrt{\frac{\alpha ^{\prime}}{2}}k_{\mu}\Delta \widetilde{W}^{\mu}(k)&=0,&\\
\Delta \widetilde{M}(k)&=0, &\\
\bigg( \frac{\alpha ^{\prime}k^{2}}{4}-1\bigg)\Delta \widetilde{W}^{\mu}(k)- \sqrt{\frac{\alpha ^{\prime}}{2}}k^{\mu}\Delta \widetilde{M}(k)&=0.&
 \end{split}
\end{eqnarray}
We choose these conditions when $k^2 = 0$ and in this case we find
\begin{equation}
\Delta \widetilde{M}(k)=0 \,, \quad \Delta \widetilde{W}^{\mu}(k) = 0 \,.
\end{equation}
Under the conditions
$\Delta\widetilde{M}(k)=0$ and $\Delta \widetilde{W}^{\mu}(k)=0$
the equations~\eqref{8} simplify as follows:
\begin{eqnarray}
\begin{split}
 \frac{\alpha ^{\prime}k^{2}}{4}\Delta \widetilde{N}(k)&=0,&\\
 \bigg( \frac{\alpha ^{\prime}k^{2}}{4}-1\bigg) \Delta \widetilde{N}(k)+ \sqrt{\frac{\alpha ^{\prime}}{2}}k_{\mu}\Delta \widetilde{V}^{\mu}(k)&=0, &\\
\frac{\alpha ^{\prime}k^{2}}{4}\Delta \widetilde{V}^{\mu}(k)&=0.&
 \end{split}
\end{eqnarray}
We choose these conditions when $k^2 \ne 0$ and in this case we find
\begin{equation}
\Delta \widetilde{N}(k)=0 \,, \quad \Delta \widetilde{V}^{\mu}(k) = 0 \,.
\end{equation}
We thus conclude that all the solutions to the equations~\eqref{8}
are equivalent to
\begin{equation}
\Delta \widetilde{M}(k)=0 \,, \quad \Delta \widetilde{N}(k)=0 \,, \quad
\Delta \widetilde{V}^{\mu}(k) = 0 \,, \quad \Delta \widetilde{W}^{\mu}(k) = 0
\end{equation}
under the gauge transformation~\eqref{gauge-transformations-for-fluctuations}.

\subsubsection{Confirmation of no additional conditions}
\label{subsection-3.2.3}

We now show that the equations
\begin{eqnarray}
\begin{split}
\frac{\alpha ^{\prime}k^{2}}{4}\widetilde{N}_{\ast}(k)-\sqrt{\frac{\alpha ^{\prime}}{2}}k_{\mu}\widetilde{W}_{\ast}^{\mu}(k)&=\mathcal{J}_{K}(k),\\
\bigg( \frac{\alpha ^{\prime}k^{2}}{4}-1\bigg) \widetilde{N}_{\ast}(k)-2\widetilde{M}_{\ast}(k)+\sqrt{\frac{\alpha ^{\prime}}{2}}k_{\mu}\widetilde{V}_{\ast}^{\mu}(k)&=\mathcal{J}_{L}(k),\\
\bigg( \frac{\alpha ^{\prime}k^{2}}{4}-1\bigg) \widetilde{W}^{\mu}_{\ast}(k)-\sqrt{\frac{\alpha ^{\prime}}{2}}k^{\mu}\widetilde{M}_{\ast}(k) +\frac{\alpha ^{\prime}k^{2}}{4}\widetilde{V}_{\ast}^{\mu}(k) &=\mathcal{J}_{U}^{\mu}(k)\label{CDE}
\end{split}
\end{eqnarray}
can be solved for $\widetilde{M}_{\ast}(k)$, $\widetilde{N}_{\ast}(k)$, $\widetilde{V}^{\mu}_{\ast}(k)$ and $\widetilde{W}^{\mu}_{\ast}(k)$
when the source terms $\mathcal{J}_{K}(k)$, $\mathcal{J}_{L}(k)$,
and $\mathcal{J}_{U}^{\mu}(k)$
are constructed from $\Psi$ satisfying
\begin{equation}
Q_{B}\Psi+\sum_{n=2}^{\infty}\frac{g^{n-1}}{n!} B \, [ \, \Psi ^{n} \, ] = 0 \,.
\label{Psi-equation-4}
\end{equation}
The equations~\eqref{CDE} correspond to
\begin{equation}
Q_{B}\widetilde{\Psi} = -\sum_{n=2}^{\infty}\frac{g^{n-1}}{n!}[ \, \Psi ^{n} \, ]
\end{equation}
at the level~$(0,1)$,
and we can show that the right-hand side of this equation
is annihilated by the BRST operator
using the equation~\eqref{Psi-equation-4} and the relations in~\eqref{h}.
At the level~$(0,1)$, this yields the following relation
among the source terms:
\begin{eqnarray}
\begin{split}
\bigg( \frac{\alpha ^{\prime}k^{2}}{4}-1\bigg) \mathcal{J}_{K}(k)- \frac{\alpha ^{\prime}k^{2}}{4}\mathcal{J}_{L}(k)+\sqrt{\frac{\alpha ^{\prime}}{2}}k^{\mu}\mathcal{J}_{U}^{\mu}(k)=0 \,. \label{CDE relation}
\end{split}
\end{eqnarray}

As in subsection~\ref{subsection-3.2.2},
we choose the conditions
\begin{equation}
\widetilde{N}_\ast (k)=0 \,, \quad
\widetilde{V}^{\mu}_\ast (k)=0 \quad \mathrm{for} \quad k^{2}=0
\end{equation}
and
\begin{equation}
\widetilde{M}_\ast (k)=0 \,, \quad
\widetilde{W}^{\mu}_\ast (k)=0\quad \mathrm{for} \quad k^{2}\ne 0
\end{equation}
to fix a gauge.
Under the conditions
$\widetilde{N}_\ast (k)=0$ and $\widetilde{V}^{\mu}_\ast (k)=0$
the equations~\eqref{CDE} simplify as follows:
\begin{align}
{}-\sqrt{\frac{\alpha ^{\prime}}{2}}k_{\mu}\widetilde{W}_{\ast}^{\mu}(k)&=\mathcal{J}_{K}(k)\,,\label{10000}\\
{}-2\widetilde{M}_{\ast}(k)&=\mathcal{J}_{L}(k)\,,
\label{10001}\\
\bigg( \frac{\alpha ^{\prime}k^{2}}{4}-1\bigg) \widetilde{W}^{\mu}_{\ast}(k)-\sqrt{\frac{\alpha ^{\prime}}{2}}k^{\mu}\widetilde{M}_{\ast}(k) &=\mathcal{J}_{U}^{\mu}(k) \,.
\label{10002}
\end{align}
We choose these conditions when $k^2 = 0$,
and in this case $\widetilde{M}_{\ast}(k)$ and $\widetilde{W}^{\mu}_{\ast}(k)$
can be solved from \eqref{10001} and \eqref{10002} as
\begin{align}
\widetilde{M}_{\ast}(k)&=-\frac{1}{2}\mathcal{J}_{L}(k),\label{10003}\\
\widetilde{W}^{\mu}_{\ast}(k)&=-\mathcal{J}^{\mu}_{U}(k)+\frac{1}{2}\sqrt{\frac{\alpha ^{\prime}}{2}}k^{\mu}\mathcal{J}_{L}(k)\label{10004}.
\end{align}
The remaining equation~\eqref{10000} is also satisfied
for $\widetilde{W}^{\mu}_{\ast}(k)$ in~\eqref{10004}
because of the relation~\eqref{CDE relation}.
Under the conditions
$\widetilde{M}_\ast (k)=0$ and $\widetilde{W}^{\mu}_\ast (k)=0$
the equations~\eqref{CDE} simplify as follows:
\begin{align}
\frac{\alpha ^{\prime}k^{2}}{4}\widetilde{N}_{\ast}(k)&=\mathcal{J}_{K}(k)\,,
\label{10006}\\
\bigg( \frac{\alpha ^{\prime}k^{2}}{4}-1\bigg) \widetilde{N}_{\ast}(k)
+\sqrt{\frac{\alpha ^{\prime}}{2}}k_{\mu}\widetilde{V}_{\ast}^{\mu}(k)&=\mathcal{J}_{L}(k)\,,
\label{10007}\\
\frac{\alpha ^{\prime}k^{2}}{4}\widetilde{V}_{\ast}^{\mu}(k) &=\mathcal{J}_{U}^{\mu}(k) \,.\label{10008}
\end{align}
We choose these conditions when $k^2 \ne 0$, and in this case we find
$\widetilde{N}_{\ast}(k)$ and $\widetilde{V}^{\mu}_{\ast}(k)$
can be solved from \eqref{10006} and \eqref{10008} as
\begin{align}
\widetilde{N}_{\ast}(k)&=\frac{4}{\alpha ^{\prime}k^{2}}\mathcal{J}_{K}(k)\, ,\label{10009}\\
\widetilde{V}_{\ast}^{\mu}(k) &=\frac{4}{\alpha ^{\prime}k^{2}}\mathcal{J}_{U}^{\mu}(k) \,.\label{10010}
\end{align}
The remaining equation~\eqref{10007} is also satisfied
for $\widetilde{N}_{\ast}(k)$ in~\eqref{10009}
and $\widetilde{V}^{\mu}_{\ast}(k)$ in~\eqref{10010}
because of the relation~\eqref{CDE relation}.

To summarize,
we have demonstrated that the equations~\eqref{CDE} can be solved
for $\widetilde{M}_{\ast}(k)$, $\widetilde{N}_{\ast}(k)$, $\widetilde{V}^{\mu}_{\ast}(k)$ and $\widetilde{W}^{\mu}_{\ast}(k)$
without imposing additional conditions on component fields of $\Psi$
satisfying the equation of motion.
We have also seen in subsection~\ref{subsection-3.2.2}
that under the gauge transformations
the component fields of $\Psi$ satisfying the equations of motion are
\begin{equation}
K(k) = 0 \,, \qquad L(k) = 0 \,, \qquad U_\mu (k) = 0
\end{equation}
and the fluctuations of $\widetilde{\Psi}$ satisfying the equations of motion are
\begin{equation}
\Delta \widetilde{M}(k)=0 \,, \quad \Delta \widetilde{N}(k)=0 \,, \quad
\Delta \widetilde{V}^{\mu}(k) = 0 \,, \quad \Delta \widetilde{W}^{\mu}(k) = 0 \,.
\end{equation}
We conclude that there are no physical excitations from the component fields
at the level~$(0,1)$.

\subsection{Massless fields}
\label{(1,1)}

Finally, let us consider the level $(1,1)$ corresponding to massless fields.
While all states at this level are annihilated by $L_0^-$,
some of them are not annihilated by $b_0^-$.
Therefore, all the component fields
of closed bosonic string field theory with the level matching condition
at the level $(1,1)$
are contained 
in closed bosonic string field theory without the level matching condition,
but additional component fields coexist at this level.
This is a new feature of the discussion at the level $(1,1)$.

Since there are many component fields at the level $(1,1)$,
we consider a subset of them.
Let us decompose string fields based on the world-sheet parity.
For $\Psi_{(1,1)}$ and $\widetilde{\Psi}_{(1,1)}$, we decompose them as follows:
\begin{equation}
\Psi_{(1,1)} = \Psi_{(1,1)}^{\rm \, odd} +\Psi_{(1,1)}^{\rm \, even} \,, \qquad
\widetilde{\Psi}_{(1,1)}
= \widetilde{\Psi}_{(1,1)}^{\rm \, odd} +\widetilde{\Psi}_{(1,1)}^{\rm \, even} \,,
\end{equation}
where the states in $\Psi_{(1,1)}^{\rm \, odd}$ and $\widetilde{\Psi}_{(1,1)}^{\rm \, odd}$
are odd under the world-sheet parity transformation
and the states in $\Psi_{(1,1)}^{\rm \, even}$ and $\widetilde{\Psi}_{(1,1)}^{\rm \, even}$
are even under the world-sheet parity transformation.
Then the kinetic terms are decomposed as
\begin{equation}
\begin{split}
& \frac{1}{2}\braket{\widetilde{\Psi}_{(1,1)},Q_{B}B\widetilde{\Psi}_{(1,1)}}
+\braket{\widetilde{\Psi}_{(1,1)},Q_{B}\Psi_{(1,1)}} \\
& = \frac{1}{2}\braket{\widetilde{\Psi}_{(1,1)}^{\rm \, even},
Q_{B}B\widetilde{\Psi}_{(1,1)}^{\rm \, even}}
+\braket{\widetilde{\Psi}_{(1,1)}^{\rm \, even},
Q_{B}\Psi_{(1,1)}^{\rm \, odd}} \\
& \quad~+\frac{1}{2}\braket{\widetilde{\Psi}_{(1,1)}^{\rm \, odd},
Q_{B}B\widetilde{\Psi}_{(1,1)}^{\rm \, odd}}
+\braket{\widetilde{\Psi}_{(1,1)}^{\rm \, odd},
Q_{B}\Psi_{(1,1)}^{\rm \, even}} \,,
\end{split}
\end{equation}
where the first two terms on the right-hand side
only contain $\Psi_{(1,1)}^{\rm \, odd}$ and $\widetilde{\Psi}_{(1,1)}^{\rm \, even}$
and the last two terms on the right-hand side
only contain $\Psi_{(1,1)}^{\rm \, even}$ and $\widetilde{\Psi}_{(1,1)}^{\rm \, odd}$.
Since the graviton is in $\Psi_{(1,1)}^{\rm \, odd}$,
we are more interested in the sector involving
$\Psi_{(1,1)}^{\rm \, odd}$ and $\widetilde{\Psi}_{(1,1)}^{\rm \, even}$.
In the rest of this subsection we will focus on this sector.

\subsubsection{The expansions in terms of component fields}

The string field $\Psi_{(1,1)}^{\mathrm{odd}}$ carrying ghost number $2$ is expanded as
\begin{equation}
\begin{split}
\Psi _{(1,1)}^{\mathrm{odd}} = \int \frac{d^{26}k}{(2\pi )^{26}}\bigg[ & B(k)c_{0}\widetilde{c}_{0}\ket{0;k}+\frac{1}{2}D(k)\big( c_{-1}c_{1}-\widetilde{c}_{-1}\widetilde{c}_{1}\big)\ket{0;k}\\
& +A_{\mu}(k)  \big( \alpha ^{\mu}_{-1}c_{0}^{-}c_{1}+\widetilde{\alpha} ^{\mu}_{-1}c_{0}^{-}\widetilde{c}_{1}\big) \ket{0;k}\\
& +E_{\mu}(k) \big( \alpha ^{\mu}_{-1}c^{+}_{0}c_{1}-\widetilde{\alpha} ^{\mu}_{-1}c_{0}^{+}\widetilde{c}_{1}\big) \ket{0;k}\\
&+\frac{1}{4}G_{\mu \nu}(k)\big( \alpha ^{\mu}_{-1}\widetilde{\alpha} ^{\nu}_{-1}+\alpha ^{\nu}_{-1}\widetilde{\alpha} ^{\mu}_{-1}\big) c_{1}\widetilde{c}_{1}\ket{0;k} \bigg] \, ,
\end{split}
\label{Psi_(1,1)-expansion}
\end{equation}
where $B(k)$, $D(k)$, $A_{\mu}(k)$, $E_{\mu}(k) $, and $G_{\mu \nu}(k)$ are five component fields.
The component field $G_{\mu \nu}(k)$ is a symmetric tensor field:
\begin{equation}
G_{\mu \nu}(k) = G_{\nu \mu}(k) \,.
\end{equation}
We can also decompose the gauge parameter $\Lambda_{(1,1)}$ as follows:
\begin{equation}
\Lambda_{(1,1)} = \Lambda_{(1,1)}^{\, \rm{odd}} + \Lambda_{(1,1)}^{\, \rm{even}}\,,
\end{equation}
where the states in $\Lambda_{(1,1)}^{\, \rm{odd}}$ are odd under the world-sheet parity transformation and the states in $\Lambda_{(1,1)}^{\, \rm{even}}$ are even under the world-sheet parity transformation.
Since the BRST operator is even under the world-sheet parity transformation,
the gauge transformation relevant for $\Psi_{(1,1)}^{\rm{odd}}$ is
\begin{equation}
\delta _{\Lambda}\Psi_{(1,1)}^{\rm{odd}} =Q_{B}\Lambda _{(1,1)}^{\rm{odd}} \,.
\label{Psi-(1,1)-odd-gauge-transformation}
\end{equation}
The gauge parameter $\Lambda_{(1,1)}^{\, \rm{odd}}$ is expanded as
\begin{equation}
\Lambda_{(1,1)}^{\, \rm{odd}}=\int \frac{d^{26}k}{(2\pi )^{26}}\ \bigg[ \chi (k)c_{0}^{-}\ket{0;k}-\frac{1}{2}\xi _{\mu}(k) \big( \alpha ^{\mu}_{-1}c_{1}-\widetilde{\alpha} ^{\mu}_{-1}\widetilde{c}_{1}\big) \ket{0;k}\bigg] \,,
\label{Lambda_(1,1)-expansion}
\end{equation}
where $\chi(k)$ and $\xi _{\mu}(k)$ are two component fields,
and the gauge transformation \eqref{Psi-(1,1)-odd-gauge-transformation} is expanded in terms of component fields as
\begin{equation}
\begin{split}
\delta _{\Lambda}B(k)&=-\frac{\alpha ^{\prime}k^{2}}{4}\chi(k)\,,\qquad 
\delta _{\Lambda}D(k)=2\chi(k)+\sqrt{\frac{\alpha ^{\prime}}{2}}k^{\mu}\xi _{\mu}(k)\,, \\ 
\delta _{\Lambda}A_{\mu}(k)&=-\sqrt{\frac{\alpha ^{\prime}}{2}}k_{\mu}\chi(k)\,,\qquad 
\delta _{\Lambda}E_{\mu}(k)=-\frac{\alpha ^{\prime}k^{2}}{4}\xi _{\mu}(k)\,, \\  
\delta _{\Lambda}G_{\mu \nu}(k)&=\sqrt{\frac{\alpha ^{\prime}}{2}}k_{\mu}\xi _{\nu}(k) +\sqrt{\frac{\alpha ^{\prime}}{2}}k_{\nu}\xi _{\mu}(k)\,.
\end{split}
\label{gauge-transformation-odd-level(1,1)-g.n.2}
\end{equation}

The string field $\widetilde{\Psi}_{(1,1)}^{\, \rm{even}}$  carrying ghost number $3$ is expanded as
\begin{equation}
\begin{split}
 \widetilde{\Psi} _{(1,1)}^{\, \rm{even}}=\int \frac{d^{26}k}{(2\pi )^{26}}\ \bigg[ & \frac{1}{2} \widetilde{C}(k)c_{0}^{+}\big( c_{-1}c_{1}+\widetilde{c}_{-1}\widetilde{c}_{1}\big) \ket{0;k}-\frac{1}{2} \widetilde{D}(k)c_{0}^{-}\big( c_{-1}c_{1}-\widetilde{c}_{-1}\widetilde{c}_{1}\big) \ket{0;k}\\
& + \frac{1}{2}\widetilde{E}_{\mu}(k)\big( \alpha _{-1}^{\mu}c_{0}c_{1}\widetilde{c}_{0}+\widetilde{\alpha }_{-1}^{\mu}c_{0}\widetilde{c}_{0}\widetilde{c}_{1}\big) \ket{0;k}\\
&+\frac{1}{2}\widetilde{F}_{\mu}(k) \big( \alpha _{-1}^{\mu}c_{1}\widetilde{c}_{-1}\widetilde{c}_{1}+\widetilde{\alpha }_{-1}^{\mu}c_{-1}c_{1}\widetilde{c}_{1}\big) \ket{0;k} \\
& -\frac{1}{4} \widetilde{G}_{\mu \nu}(k)\big( \alpha _{-1}^{\mu}\widetilde{\alpha} _{-1}^{\nu}+\alpha _{-1}^{\nu}\widetilde{\alpha} _{-1}^{\mu}\big) c_{0}^{-}c_{1}\widetilde{c}_{1}\ket{0;k}\\
& +\frac{1}{4} \widetilde{H}_{\mu \nu}(k)\big( \alpha _{-1}^{\mu}\widetilde{\alpha} _{-1}^{\nu}-\alpha _{-1}^{\nu}\widetilde{\alpha} _{-1}^{\mu}\big) c_{0}^{+}c_{1}\widetilde{c}_{1}\ket{0;k}\bigg] \, ,
\end{split}
\label{Psi-tilde_(1,1)-expansion}
\end{equation}
where $\widetilde{C}(k)$, $\widetilde{D}(k)$, $\widetilde{E}_{\mu}(k)$, $\widetilde{F}_{\mu}(k)$, $\widetilde{G}_{\mu \nu}(k)$, and $\widetilde{H}_{\mu \nu}(k)$ are six component fields.
The component field $\widetilde{G}_{\mu \nu}(k)$ is a symmetric tensor field
and the component field $\widetilde{H}_{\mu \nu}(k)$ is an antisymmetric tensor field:
\begin{equation}
\widetilde{G}_{\mu \nu}(k) = \widetilde{G}_{\nu \mu}(k) \,, \qquad
\widetilde{H}_{\mu \nu}(k) = {}-\widetilde{H}_{\nu \mu}(k) \,.
\end{equation}
We again decompose the gauge parameter $\widetilde{\Lambda}_{(1,1)}$ as follows:
\begin{equation}
\widetilde{\Lambda}_{(1,1)}
= \widetilde{\Lambda}_{(1,1)}^{\, \rm{odd}} +\widetilde{\Lambda}^{\, \rm{even}}_{(1,1)} \,,
\end{equation}
where the states in $\widetilde{\Lambda}_{(1,1)}^{\, \rm{odd}}$
are odd under the world-sheet parity transformation
and the states in $\widetilde{\Lambda}_{(1,1)}^{\, \rm{even}}$
are even under the world-sheet parity transformation.
The gauge transformation relevant for $\widetilde{\Psi}_{(1,1)}^{\rm{even}}$ is
\begin{equation}
\delta _{\Lambda}\widetilde{\Psi}_{(1,1)}^{\rm{even}} =Q_{B}\widetilde{\Lambda}_{(1,1)}^{\rm{even}} \,, \label{Psi-tilde-(1,1)-even-gauge-transformation}
\end{equation}
and the gauge parameter $\widetilde{\Lambda }_{(1,1)}^{\, \rm{even}}$ is expanded as
\begin{eqnarray}
\begin{split}
\widetilde{\Lambda }_{(1,1)}^{\, \rm{even}}=\int \frac{d^{26}k}{(2\pi )^{26}}\ \bigg[ &\frac{1}{4}\widetilde{\eta}(k)(c_{-1}c_{1}+\widetilde{c}_{-1}\widetilde{c}_{1})\ket{0;k}+\frac{1}{2}\widetilde{\zeta}_{\mu}(k)(\alpha ^{\mu}_{-1}c_{0}^{+}c_{1}+\widetilde{\alpha} ^{\mu}_{-1}c_{0}^{+}\widetilde{c}_{1})\ket{0;k}& \\
&+\frac{1}{2}\widetilde{\xi}_{\mu}(k)(\alpha ^{\mu}_{-1}c_{0}^{-}c_{1}-\widetilde{\alpha} ^{\mu}_{-1}c_{0}^{-}\widetilde{c}_{1})\ket{0;k}&\\
&+\frac{1}{8}\widetilde{\omega}_{\mu \nu}(k)(\alpha ^{\mu}_{-1}\widetilde{\alpha} ^{\nu}_{-1}-\alpha ^{\nu}_{-1}\widetilde{\alpha} ^{\mu}_{-1})c_{1}\widetilde{c}_{1}\ket{0;k}\bigg] \,,
\end{split}
\end{eqnarray}
where $\widetilde{\eta}(k)$, $\widetilde{\zeta}_{\mu}(k)$, $\widetilde{\xi}_{\mu}(k)$, and $\widetilde{\omega}_{\mu \nu}(k)$ are four component fields.
The gauge transformation \eqref{Psi-tilde-(1,1)-even-gauge-transformation} is expanded in terms of component fields as
\begin{eqnarray}
\begin{split}
\delta _{\Lambda}\widetilde{C}(k)&=\frac{\alpha ^{\prime}k^{2}}{4} \widetilde{\eta}(k) -\sqrt{\frac{\alpha ^{\prime}}{2}}k^{\mu}\widetilde{\zeta}_{\mu}(k) \,, \\
\delta _{\Lambda}\widetilde{D}(k)&=\sqrt{\frac{\alpha ^{\prime}}{2}}k^{\mu}\widetilde{\xi}_{\mu}(k) \,, \quad
\delta _{\Lambda}\widetilde{E}_{\mu}(k)=\frac{\alpha ^{\prime}k^{2}}{4}\widetilde{\xi}_{\mu}(k) \,, \\
\delta _{\Lambda}\widetilde{F}_{\mu}(k)&=\frac{1}{2}\sqrt{\frac{\alpha ^{\prime}}{2}}k_{\mu}\widetilde{\eta}(k)-\widetilde{\zeta}_{\mu}(k) +\widetilde{\xi}_{\mu}(k) -\frac{1}{2}\sqrt{\frac{\alpha ^{\prime}}{2}}k^{\nu} \widetilde{\omega}_{\mu \nu}(k) \,, \\
\delta _{\Lambda}\widetilde{G}_{\mu \nu}(k)&=-\sqrt{\frac{\alpha ^{\prime}}{2}}k_{\mu}\widetilde{\xi}_{\nu}(k)  -\sqrt{\frac{\alpha ^{\prime}}{2}}k_{\nu}\widetilde{\xi}_{\mu}(k)\,,\\
\delta _{\Lambda}\widetilde{H}_{\mu \nu}(k)&=-\sqrt{\frac{\alpha ^{\prime}}{2}}k_{\mu} \widetilde{\zeta}_{\nu}(k) +\sqrt{\frac{\alpha ^{\prime}}{2}}k_{\nu} \widetilde{\zeta}_{\mu}(k) +\frac{\alpha ^{\prime}k^{2}}{4}\widetilde{\omega}_{\mu \nu}(k) \, .
\end{split}
\label{gauge-even-transformation-level(1,1)-g.n.3}
\end{eqnarray}
Note that gauge transformations parameterized
by $\widetilde{\eta}(k)$, $\widetilde{\zeta}_{\mu}(k)$, $\widetilde{\xi}_{\mu}(k)$, and $\widetilde{\omega}_{\mu \nu}(k)$
are not independent
and the gauge transformation of the form
\begin{equation}
\widetilde{\Lambda }_{(1,1)}^{\, \rm{even}} = Q_B \widetilde{\Omega }_{(1,1)}^{\, \rm{even}}
\end{equation}
with
\begin{equation}
\widetilde{\Omega} _{(1,1)}^{\, \rm{even}}=\int \frac{d^{26}k}{(2\pi )^{26}}\ \bigg[ \widetilde{\varepsilon} (k)c_{0}^{+}\ket{0;k}+\frac{1}{2}\widetilde{\pi} _{\mu}(k) \big( \alpha ^{\mu}_{-1}c_{1}+\widetilde{\alpha} ^{\mu}_{-1}\widetilde{c}_{1}\big) \ket{0;k}\bigg] \,,
\end{equation}
where $\widetilde{\varepsilon} (k)$ and $\widetilde{\pi} _{\mu}(k)$ are component fields,
does not change $\widetilde{\Psi}_{(1,1)}^{\, \rm{even}}$ because $Q_B^2 = 0$.
In terms of component fields, this can be expressed as
\begin{equation}
\begin{split}
\widetilde{\eta}(k)&={}-4\widetilde{\varepsilon} (k)+2\sqrt{\frac{\alpha ^{\prime}}{2}}k^{\mu}\widetilde{\pi}_{\mu}(k)\,,\\
\widetilde{\zeta}_{\mu}(k)&={}-2\sqrt{\frac{\alpha ^{\prime}}{2}}k_{\mu}\widetilde{\varepsilon} (k)+\frac{\alpha ^{\prime}k^{2}}{2}\widetilde{\pi} _{\mu}(k)\,, \\ 
\widetilde{\omega}_{\mu \nu}(k)&=2\sqrt{\frac{\alpha ^{\prime}}{2}}k_{\mu}\widetilde{\pi} _{\nu}(k) -2\sqrt{\frac{\alpha ^{\prime}}{2}}k_{\nu}\widetilde{\pi} _{\mu}(k) \,.
\end{split}
\label{degree-of-freedoms-of-gauge-parameters-(1,1)}
\end{equation}

Let us next expand the kinetic terms.
We find
\begin{eqnarray}
\begin{split}
 \frac{1}{2}\braket{\widetilde{\Psi}_{(1,1)}^{\rm \, even},Q_{B}B\widetilde{\Psi}_{(1,1)}^{\rm \, even}}&\\
 = \frac{1}{2}\int \frac{d^{26}k}{(2\pi )^{26}}\ \bigg( &\frac{1}{2}\frac{\alpha ^{\prime}k^{2}}{4}\widetilde{D}(-k)\widetilde{D}(k)
 +\sqrt{\frac{\alpha ^{\prime}}{2}}k_{\mu}\widetilde{E}^{\mu}(-k)\widetilde{D}(k)
 +\widetilde{E}^{\mu}(-k)\widetilde{E}_{\mu}(k)  \\
&-\frac{1}{2}\sqrt{\frac{\alpha ^{\prime}}{2}}\widetilde{G}^{\mu \nu}(-k)\bigg[ k_{\mu}\widetilde{E}_{\nu}(k)+k_{\nu}\widetilde{E}_{\mu}(k)\bigg]
-\frac{1}{4}\frac{\alpha ^{\prime}k^{2}}{4}\widetilde{G}^{\mu \nu}(-k)\widetilde{G}_{\mu \nu}(k)
  \bigg) ,
\end{split}
\end{eqnarray}
and
\begin{eqnarray}
\begin{split}
\braket{\widetilde{\Psi}_{(1,1)}^{\rm \, even},Q_{B}\Psi_{(1,1)}^{\rm \, odd}}=&\int \frac{d^{26}k}{(2\pi )^{26}}\ \bigg( \frac{1}{2}\widetilde{D}(-k) \bigg[ -\frac{\alpha ^{\prime}k^{2}}{4}D(k)+2B(k)+\sqrt{\frac{\alpha ^{\prime}}{2}}k^{\mu}E_{\mu}(k)\bigg] \\
&+\frac{1}{2}\widetilde{E}^{\mu}(-k) \bigg[ -\sqrt{\frac{\alpha ^{\prime}}{2}}k_{\mu}D(k)+2A_{\mu}(k)-2E_{\mu}(k)-\sqrt{\frac{\alpha ^{\prime}}{2}}k^{\nu}G_{\mu \nu}(k)\bigg]  \\
&+\frac{1}{4}\widetilde{G}^{\mu \nu}(-k)\bigg[ \sqrt{\frac{\alpha ^{\prime}}{2}}k_{\mu}E_{\nu}(k) +\sqrt{\frac{\alpha ^{\prime}}{2}}k_{\nu} E_{\mu}(k)+\frac{\alpha ^{\prime}k^{2}}{4}G_{\mu \nu}(k)   \bigg]  \\
&+\frac{1}{2}\widetilde{C}(-k) \bigg[ 2B(k)-\sqrt{\frac{\alpha ^{\prime}}{2}}k^{\mu}A_{\mu}(k) \bigg] \\
&+\widetilde{F}^{\mu}(-k) \bigg[ \sqrt{\frac{\alpha ^{\prime}}{2}}k_{\mu}B(k)-\frac{\alpha ^{\prime}k^{2}}{4}A_{\mu}(k) \bigg]  \\
&+\frac{1}{4}\widetilde{H}^{\mu \nu}(-k)\bigg[ -\sqrt{\frac{\alpha ^{\prime}}{2}}k_{\mu}A_{\nu}(k)  +\sqrt{\frac{\alpha ^{\prime}}{2}}k_{\nu}A_{\mu}(k)  \bigg]
 \bigg) .
\end{split}
\end{eqnarray}
The equation of motion
\begin{equation}
Q_{B}B\widetilde{\Psi}+Q_{B}\Psi  = 0
\end{equation}
is expanded in terms of component fields in the sector  involving
$\Psi_{(1,1)}^{\rm \, odd}$ and $\widetilde{\Psi}_{(1,1)}^{\rm \, even}$ as
\begin{equation}
\begin{split}
&k^{\mu}A^{\nu}(k)-k^{\nu}A^{\mu}(k)=0\, ,\\
&2B(k)-\sqrt{\frac{\alpha ^{\prime}}{2}}k_{\mu}A^{\mu}(k)=0\,,  \\
&\sqrt{\frac{\alpha ^{\prime}}{2}}k^{\mu}B(k)-\frac{\alpha ^{\prime}k^{2}}{4}A^{\mu}(k)=0\,, \\
&\frac{\alpha ^{\prime}k^{2}}{4}\widetilde{D}(k)-\sqrt{\frac{\alpha ^{\prime}}{2}}k^{\mu}\widetilde{E}_{\mu}(k) =\frac{\alpha ^{\prime}k^{2}}{4}D(k)-2B(k)-\sqrt{\frac{\alpha ^{\prime}}{2}}k^{\mu}E_{\mu}(k)\, ,  \\
&\frac{1}{2}\sqrt{\frac{\alpha ^{\prime}}{2}}k^{\mu}\widetilde{D}(k)+\widetilde{E}^{\mu}(k)+\frac{1}{2}\sqrt{\frac{\alpha ^{\prime}}{2}}k_{\nu}\widetilde{G}^{\mu \nu}(k) \\
&\qquad \qquad =\frac{1}{2}\sqrt{\frac{\alpha ^{\prime}}{2}}k^{\mu}D(k)-A^{\mu}(k)+E^{\mu}(k)+\frac{1}{2}\sqrt{\frac{\alpha ^{\prime}}{2}}k_{\nu}G^{\mu \nu}(k)\,, \\
& \sqrt{\frac{\alpha ^{\prime}}{2}}\bigg( k^{\nu}\widetilde{E}^{\mu}(k)+k^{\mu}\widetilde{E}^{\nu}(k) \bigg) +\frac{\alpha ^{\prime}k^{2}}{4}\widetilde{G}^{\mu \nu}(k) \\
&\qquad  \qquad =\sqrt{\frac{\alpha ^{\prime}}{2}}\bigg( k^{\mu}E^{\nu}(k)+k^{\nu}E^{\mu}(k)\bigg) +\frac{\alpha ^{\prime}k^{2}}{4}G^{\mu \nu}(k)\,,
\end{split}
\label{comp_eom_psi_tilde}
\end{equation}
and the equation of motion
\begin{equation}
Q_{B}\widetilde{\Psi}+\sum_{n=2}^{\infty}\frac{g^{n-1}}{n!}[ \, \Psi ^{n} \, ]  = 0
\label{Psi-tilde-equation-5}
\end{equation}
is expanded in terms of component fields in this sector as
\begin{equation}
\begin{split}
&\frac{\alpha ^{\prime}k^{2}}{4}\widetilde{D}(k) -\sqrt{\frac{\alpha ^{\prime}}{2}}k^{\mu} \widetilde{E}_{\mu}(k)=\mathcal{J}_{D}(k), \\
&\frac{1}{2}\sqrt{\frac{\alpha ^{\prime}}{2}}k^{\mu} \widetilde{D}(k)+ \widetilde{E}^{\mu}(k)+\frac{1}{2}\sqrt{\frac{\alpha ^{\prime}}{2}}k_{\nu} \widetilde{G}^{\mu \nu}(k) =\mathcal{J}_{E}^{\mu}(k)\, , \\
&\frac{1}{4}\sqrt{\frac{\alpha ^{\prime}}{2}}\bigg( k^{\nu}\widetilde{E}^{\mu}(k)+k^{\mu}\widetilde{E}^{\nu}(k)\bigg)+\frac{\alpha ^{\prime}k^{2}}{16}\widetilde{G}^{\mu \nu}(k)=\mathcal{J}_{G}^{\mu \nu}(k)\, ,\\
&\widetilde{C}(k)+\widetilde{D}(k)-\sqrt{\frac{\alpha ^{\prime}}{2}}k^{\mu}\widetilde{F}_{\mu}(k)=\mathcal{J}_{B}(k)\,,  \\
&\frac{1}{2}\sqrt{\frac{\alpha ^{\prime}}{2}}k^{\mu} \widetilde{C}(k)  -\frac{\alpha ^{\prime}k^{2}}{4}\widetilde{F}^{\mu}(k)+ \widetilde{E}^{\mu}(k)-\frac{1}{2}\sqrt{\frac{\alpha ^{\prime}}{2}}k_{\nu} \widetilde{H}^{\mu \nu}(k)=\mathcal{J}_{A}^{\mu}(k) \,, 
\end{split}
\label{comp_eq_Psi_tilde_(1,1)}
\end{equation}
where the source terms $\mathcal{J}_{B}(k)$, $\mathcal{J}_{D}(k)$, $\mathcal{J}_{A}^{\mu}(k)$, $\mathcal{J}_{E}^{\mu}(k)$, and $\mathcal{J}_{G}^{\mu \nu}(k)$ are from the interaction terms
which depend only on the string field $\Psi$. 
After eliminating $\widetilde{\Psi}$,
the equation for $\Psi$~\eqref{Psi-equation},
\begin{equation}
Q_{B}\Psi+\sum_{n=2}^{\infty}\frac{g^{n-1}}{n!} B \, [ \, \Psi ^{n} \, ] = 0 \,,
\label{Psi-equation-5}
\end{equation}
is expanded as
\begin{equation}
\begin{split}
\frac{\alpha ^{\prime}k^{2}}{4}D(k)-2B(k)-\sqrt{\frac{\alpha ^{\prime}}{2}}k^{\mu}E_{\mu}(k) & =\mathcal{J}_{D}(k) \,, \\
\frac{1}{2}\sqrt{\frac{\alpha ^{\prime}}{2}}k^{\mu}D(k)-A^{\mu}(k)+E^{\mu}(k)+\frac{1}{2}\sqrt{\frac{\alpha ^{\prime}}{2}}k_{\nu}G^{\mu \nu}(k)
& = \mathcal{J}_{E}^{\mu}(k) \,, \\
\frac{1}{4}\sqrt{\frac{\alpha ^{\prime}}{2}}k^{\mu}E^{\nu}(k)  +\frac{1}{4}\sqrt{\frac{\alpha ^{\prime}}{2}}k^{\nu}E^{\mu}(k) +\frac{\alpha ^{\prime}k^{2}}{16}G^{\mu \nu}(k)
& = \mathcal{J}_{G}^{\mu \nu}(k) \,,\\
2B(k)-\sqrt{\frac{\alpha ^{\prime}}{2}}k_{\mu}A^{\mu}(k) & =0 \,, \\
\sqrt{\frac{\alpha ^{\prime}}{2}}k^{\mu}B(k)-\frac{\alpha ^{\prime}k^{2}}{4}A^{\mu}(k)
& = 0 \,,\\
k^{\mu}A^{\nu}(k)-k^{\nu}A^{\mu}(k) & = 0 \,.
\end{split}
\label{comp_eq_Psi_(1,1)}
\end{equation}

\subsubsection{The equations of motion for the interacting fields}
\label{subsection-3.3.2}

In the expansion~\eqref{Psi_(1,1)-expansion} of $\Psi_{(1,1)}^{\mathrm{odd}}$,
the states $c_{0}\widetilde{c}_{0}\ket{0;k}$
and $\big( \alpha ^{\mu}_{-1}c_{0}^{-}c_{1}+\widetilde{\alpha} ^{\mu}_{-1}c_{0}^{-}\widetilde{c}_{1}\big) \ket{0;k}$ are not annihilated by $b_0^-$.
Therefore, the corresponding component fields $B(k)$ and $A_{\mu}(k)$
are absent in closed string field theory with the level-matching condition.
On the other hand, the component fields $D(k)$, $E_{\mu}(k)$, and $G_{\mu \nu}(k)$
exist in closed string field theory with the level-matching condition
and the graviton and the dilaton are described by these fields.
The equations of motion for these fields
in closed string field theory with the level-matching condition
are given by
\begin{equation}
\begin{split}
\frac{\alpha ^{\prime}k^{2}}{4}D(k)-\sqrt{\frac{\alpha ^{\prime}}{2}}k^{\mu}E_{\mu}(k) & =\mathcal{J}_{D}(k) \,, \\
\frac{1}{2}\sqrt{\frac{\alpha ^{\prime}}{2}}k^{\mu}D(k)+E^{\mu}(k)+\frac{1}{2}\sqrt{\frac{\alpha ^{\prime}}{2}}k_{\nu}G^{\mu \nu}(k)
& = \mathcal{J}_{E}^{\mu}(k) \,, \\
\frac{1}{4}\sqrt{\frac{\alpha ^{\prime}}{2}}k^{\mu}E^{\nu}(k)  +\frac{1}{4}\sqrt{\frac{\alpha ^{\prime}}{2}}k^{\nu}E^{\mu}(k) +\frac{\alpha ^{\prime}k^{2}}{16}G^{\mu \nu}(k)
& = \mathcal{J}_{G}^{\mu \nu}(k) \,,
\end{split}
\label{comp_eq_Psi_(1,1)_level-matching}
\end{equation}
which are obtained from~\eqref{comp_eq_Psi_(1,1)}
by simply setting $B(k)=0$ and $A^{\mu}(k)=0$.

It is easy to see the equivalence of the equations~\eqref{comp_eq_Psi_(1,1)}
in closed string field theory without the level-matching condition
to the equations~\eqref{comp_eq_Psi_(1,1)_level-matching}
in closed string field theory with the level-matching condition
up to gauge transformations.
The equation
\begin{equation}
k^{\mu}A^{\nu}(k)-k^{\nu}A^{\mu}(k) = 0
\end{equation}
means that the field strength of $A^{\mu}(k)$ vanishes.
Therefore, we can bring any solution to the form where
\begin{equation}
A^{\mu}(k) = 0
\end{equation}
by the gauge transformation using the gauge parameter $\chi(k)$
in~\eqref{gauge-transformation-odd-level(1,1)-g.n.2}.
In this gauge, the component field $B(k)$ also vanishes:
\begin{equation}
B(k) = 0 \,.
\end{equation}
This establishes the equivalence of~\eqref{comp_eq_Psi_(1,1)}
to~\eqref{comp_eq_Psi_(1,1)_level-matching} under the gauge transformation.

Note that we can have solutions which cannot be brought
to the form where $A^{\mu}(k)=0$ by the gauge transformation
if we compactify the target space on a torus.
In this case closed string field theory without the level-matching condition
can be inequivalent
to closed string field theory with the level-matching condition
nonperturbatively.

\subsubsection{The equations of motion for the fields in the additional string field}
\label{subsection-3.3.3}

Let us next consider the equations~\eqref{comp_eq_Psi_tilde_(1,1)}.
The source terms $\mathcal{J}_{B}(k)$, $\mathcal{J}_{D}(k)$, $\mathcal{J}^{\mu}_{A}(k)$, $\mathcal{J}^{\mu}_{E}(k)$, and $\mathcal{J}^{\mu \nu}_{G}(k)$
do not depend on component fields of $\widetilde{\Psi}$,
but they depend on component fields of $\Psi$ at all levels.
Once we have a solution to
\begin{equation}
Q_{B}\Psi+\sum_{n=2}^{\infty}\frac{g^{n-1}}{n!} B \, [ \, \Psi ^{n} \, ] = 0 \,, \label{Psi-equation-6}
\end{equation}
the equations~\eqref{comp_eq_Psi_tilde_(1,1)} can be regarded as the equations
for $\widetilde{C}(k)$, $\widetilde{D}(k)$, $\widetilde{E}^{\mu}(k)$, $\widetilde{F}^{\mu}(k)$, $ \widetilde{G}^{\mu \nu}(k)$, and $\widetilde{H}^{\mu \nu}(k)$.
We will show in subsection~\ref{subsection-3.3.4} that
the equations~\eqref{comp_eq_Psi_tilde_(1,1)} can be solved
for the source terms from $\Psi$ satisfying~\eqref{Psi-equation-6},
and we denote the solutions by $\widetilde{C}_{\ast}(k)$, $\widetilde{D}_{\ast}(k)$, $\widetilde{E}^{\mu}_{\ast}(k)$, $\widetilde{F}^{\mu}_{\ast}(k)$, $ \widetilde{G}^{\mu \nu}_{\ast}(k)$, and $\widetilde{H}^{\mu \nu}_{\ast}(k)$.
As in the discussion of the level $(0,1)$,
we first consider the equations for fluctuations around the solution.
We expand $\widetilde{C}(k)$, $\widetilde{D}(k)$, $\widetilde{E}^{\mu}(k)$, $\widetilde{F}^{\mu}(k)$, $ \widetilde{G}^{\mu \nu}(k)$, and $\widetilde{H}^{\mu \nu}(k)$ as
\begin{equation}
\begin{split}
\widetilde{C}(k)&=\widetilde{C}_{\ast}(k)+\Delta \widetilde{C}(k)\, ,\\
\widetilde{D}(k)&=\widetilde{D}_{\ast}(k)+\Delta \widetilde{D}(k)\, , \\
\widetilde{E}^{\mu}(k)&=\widetilde{E}^{\mu}_{\ast}(k)+\Delta \widetilde{E}^{\mu}(k)\, , \\
\widetilde{F}^{\mu}(k)&=\widetilde{F}^{\mu}_{\ast}(k)+\Delta \widetilde{F}^{\mu}(k)\, , \\
\widetilde{G}^{\mu \nu}(k)&=\widetilde{G}^{\mu \nu}_{\ast}(k)+\Delta \widetilde{G}^{\mu \nu}(k)\, , \\
\widetilde{H}^{\mu \nu}(k)&=\widetilde{H}^{\mu\nu}_{\ast}(k)+\Delta \widetilde{H}^{\mu\nu}(k)\, , 
\end{split}
\end{equation}
and the equations for the fluctuations $\Delta \widetilde{C}(k)$, $\Delta \widetilde{D}(k)$, $\Delta \widetilde{E}^{\mu}(k)$, $\Delta \widetilde{F}^{\mu}(k)$, $\Delta \widetilde{G}^{\mu \nu}(k)$, and $\Delta \widetilde{H}^{\mu \nu}(k)$ are given by
\begin{eqnarray}
\begin{split}
&\frac{\alpha ^{\prime}k^{2}}{4}\Delta \widetilde{D}(k) -\sqrt{\frac{\alpha ^{\prime}}{2}}k^{\mu}\Delta \widetilde{E}_{\mu}(k)=0\,, \\
&\frac{1}{2}\sqrt{\frac{\alpha ^{\prime}}{2}}k^{\mu}\Delta \widetilde{D}(k)+\Delta \widetilde{E}^{\mu}(k)+\frac{1}{2}\sqrt{\frac{\alpha ^{\prime}}{2}}k_{\nu}\Delta \widetilde{G}^{\mu \nu}(k) =0\, , \\
&\frac{1}{4}\sqrt{\frac{\alpha ^{\prime}}{2}}\bigg( k^{\nu}\Delta \widetilde{E}^{\mu}(k)+k^{\mu}\Delta \widetilde{E}^{\nu}(k)\bigg)+\frac{\alpha ^{\prime}k^{2}}{16}\Delta \widetilde{G}^{\mu \nu}(k)=0\,, \\
&\Delta \widetilde{C}(k)+\Delta \widetilde{D}(k)-\sqrt{\frac{\alpha ^{\prime}}{2}}k^{\mu} \Delta \widetilde{F}_{\mu}(k)=0\,,  \\
&\frac{1}{2}\sqrt{\frac{\alpha ^{\prime}}{2}}k^{\mu}\Delta \widetilde{C}(k)  -\frac{\alpha ^{\prime}k^{2}}{4}\Delta \widetilde{F}^{\mu}(k)+\Delta \widetilde{E}^{\mu}(k)-\frac{1}{2}\sqrt{\frac{\alpha ^{\prime}}{2}}k_{\nu}\Delta \widetilde{H}^{\mu \nu}(k)=0 \,.
\end{split}
\label{Psi-tilde-fluctuation-level-(1,1)}
\end{eqnarray}

In the expansion~\eqref{Psi-tilde_(1,1)-expansion} of the string field $\widetilde{\Psi}_{(1,1)}^{\mathrm{even}}$,
the three states $c_{0}^{+}\big( c_{-1}c_{1}+\widetilde{c}_{-1}\widetilde{c}_{1}\big) \ket{0;k}$,
$\big( \alpha _{-1}^{\mu}c_{1}\widetilde{c}_{-1}\widetilde{c}_{1}+\widetilde{\alpha }_{-1}^{\mu}c_{-1}c_{1}\widetilde{c}_{1}\big) \ket{0;k}$,
and $\big( \alpha _{-1}^{\mu}\widetilde{\alpha} _{-1}^{\nu}-\alpha _{-1}^{\nu}\widetilde{\alpha} _{-1}^{\mu}\big) c_{0}^{+}c_{1}\widetilde{c}_{1}\ket{0;k}$
are annihilated by $b_0^-$,
and the corresponding component fields $\Delta \widetilde{C}(k)$, $\Delta \widetilde{F}_{\mu}(k)$,
and $\Delta \widetilde{H}_{\mu \nu}(k)$
only appear in the last two equations of~\eqref{Psi-tilde-fluctuation-level-(1,1)}.
Let us first focus on the the first three equations of \eqref{Psi-tilde-fluctuation-level-(1,1)} which only involve $\Delta \widetilde{D}(k)$, $\Delta \widetilde{E}^{\mu}(k)$, and $\Delta \widetilde{G}^{\mu \nu}(k)$:
\begin{align}
&\frac{\alpha ^{\prime}k^{2}}{4}\Delta \widetilde{D}(k) -\sqrt{\frac{\alpha ^{\prime}}{2}}k^{\mu}\Delta \widetilde{E}_{\mu}(k)=0\,, \label{Psi-tilde-fluctuation-level-(1,1)-L-U-S-1}\\
&\frac{1}{2}\sqrt{\frac{\alpha ^{\prime}}{2}}k^{\mu}\Delta \widetilde{D}(k)+\Delta \widetilde{E}^{\mu}(k)+\frac{1}{2}\sqrt{\frac{\alpha ^{\prime}}{2}}k_{\nu}\Delta \widetilde{G}^{\mu \nu}(k) =0\, ,\label{Psi-tilde-fluctuation-level-(1,1)-L-U-S-2} \\
&\frac{1}{4}\sqrt{\frac{\alpha ^{\prime}}{2}}\bigg( k^{\nu}\Delta \widetilde{E}^{\mu}(k)+k^{\mu}\Delta \widetilde{E}^{\nu}(k)\bigg)+\frac{\alpha ^{\prime}k^{2}}{16}\Delta \widetilde{G}^{\mu \nu}(k)=0\,.\label{Psi-tilde-fluctuation-level-(1,1)-L-U-S-3}
\end{align}
They coincide with \eqref{comp_eq_Psi_(1,1)_level-matching}
without the source terms
by the replacement
$D(k)$, $E^{\mu}(k)$, and ${G}^{\mu \nu}(k)$
with
$\Delta \widetilde{D}(k)$, $\Delta \widetilde{E}^{\mu}(k)$,
and $\Delta \widetilde{G}^{\mu \nu}(k)$, respectively.
We thus know that they describe a copy of the system
which consists of a free dilaton and a free graviton.
The fields $\Delta \widetilde{D}(k)$ and $\Delta \widetilde{E}^{\mu}(k)$
also appear in the remaining equations of~\eqref{Psi-tilde-fluctuation-level-(1,1)},
and their expressions depend on a gauge we choose.
It is therefore convenient to fix a gauge,
and we use the light-cone gauge.
We define
\begin{equation}
k^+ = \frac{1}{\sqrt{2}} ( k^0 +k^1 ) \,,\qquad
k^- = \frac{1}{\sqrt{2}} ( k^0 -k^1 ) \,,
\end{equation}
and we work in a Lorentz frame where $k^+ \ne 0$.\footnote{
As we mentioned before, we will not discuss possible subtleties associated with zero-momentum states.
}
We solve~\eqref{Psi-tilde-fluctuation-level-(1,1)-L-U-S-2}
for $\Delta \widetilde{E}^{\mu}(k)$ as
\begin{equation}
\Delta \widetilde{E}^{\mu}(k)=-\frac{1}{2}\sqrt{\frac{\alpha ^{\prime}}{2}}k^{\mu}\Delta \widetilde{D}(k)-\frac{1}{2}\sqrt{\frac{\alpha ^{\prime}}{2}}k_{\nu}\Delta \widetilde{G}^{\mu \nu}(k) \,, \label{Psi-tilde-fluctuation-level-(1,1)-L-U-S-6}
\end{equation}
and we eliminate $\Delta \widetilde{E}^{\mu}(k)$ from
\eqref{Psi-tilde-fluctuation-level-(1,1)-L-U-S-1}
and \eqref{Psi-tilde-fluctuation-level-(1,1)-L-U-S-3} to obtain
\begin{align}
&2k^{2}\Delta \widetilde{D}(k) +k_{\mu}k_{\nu}\Delta \widetilde{G}^{\mu \nu}(k)=0\,,\label{Psi-tilde-fluctuation-level-(1,1)-L-U-S-4} \\
& 2k^{\mu}k^{\nu}\Delta \widetilde{D}(k)+k_{\rho}k^{\nu}\Delta \widetilde{G}^{\mu \rho}(k)+k_{\rho}k^{\mu}\Delta \widetilde{G}^{\nu \rho}(k)-k^{2}\Delta \widetilde{G}^{\mu \nu}(k)=0\,.\label{Psi-tilde-fluctuation-level-(1,1)-L-U-S-5}
\end{align}
We choose the condition
\begin{equation}
\Delta \widetilde{G}^{\mu +}(k) = 0
\label{G-tilde-gauge}
\end{equation}
to fix a gauge.
This condition can be satisfied by the gauge transformation~\eqref{gauge-even-transformation-level(1,1)-g.n.3}
with the parameter $\widetilde{\xi}_{\mu}(k)$ when $k^+ \ne 0$.
Under this gauge, 
the equation~\eqref{Psi-tilde-fluctuation-level-(1,1)-L-U-S-5}
with $\mu=+$ and $\nu=+$ simplifies as follows:
\begin{equation}
2 \, (k^+)^2 \Delta \widetilde{D}(k)=0\,.
\end{equation}
Since $k^+ \ne 0$, we obtain
\begin{equation}
\Delta \widetilde{D}(k)=0\,.
\end{equation}
The equation~\eqref{Psi-tilde-fluctuation-level-(1,1)-L-U-S-5}
with $\mu=+$ is now given by
\begin{equation}
k^+ k_{\rho}\Delta \widetilde{G}^{\nu \rho}(k)=0\,.
\end{equation}
Since $k^+ \ne 0$, we obtain
\begin{equation}
k_{\nu}\Delta \widetilde{G}^{\mu \nu}(k) = 0 \,.
\label{k_nu-G-tilde-mu-nu}
\end{equation}
When this equation is satisfied,
the equation~\eqref{Psi-tilde-fluctuation-level-(1,1)-L-U-S-4}
is also satisfied,
and we find that $\Delta \widetilde{E}^{\mu}(k)$ vanishes:
\begin{equation}
\Delta \widetilde{E}^{\mu}(k) = 0 \,.
\end{equation}
The equation~\eqref{Psi-tilde-fluctuation-level-(1,1)-L-U-S-5}
further simplifies as follows:
\begin{equation}
k^{2}\Delta \widetilde{G}^{\mu \nu}(k)=0\, .\label{Psi-tilde-fluctuation-level-(1,1)-L-U-S-15}
\end{equation}
We thus conclude $\Delta \widetilde{G}^{\mu \nu}(k)=0$ when $k^2 \ne 0$.
When $k^2 = 0$,
we need to solve the equation~\eqref{k_nu-G-tilde-mu-nu}.
Under the condition~\eqref{G-tilde-gauge},
the remaining components of $\Delta \widetilde{G}^{\mu \nu}(k)$ are
\begin{equation}
(\Delta \widetilde{G}^{IJ}(k), \, \Delta \widetilde{G}^{I-}(k), \, \Delta \widetilde{G}^{--}(k))\,.
\end{equation}
Here and in what follows we use $I$ and $J$ to label transverse directions.
The equation~\eqref{k_nu-G-tilde-mu-nu} with $\mu = +$
is satisfied by our gauge choice~\eqref{G-tilde-gauge}.
The equation~\eqref{k_nu-G-tilde-mu-nu} with $\mu = I$ is given by
\begin{equation}
-k^{+}\Delta \widetilde{G}^{I -}(k)+k_{J}\Delta \widetilde{G}^{I J}(k)=0 \,.
\end{equation}
This is satisfied by choosing $\Delta \widetilde{G}^{I -}(k)$ as
\begin{equation}
\Delta \widetilde{G}^{I -}(k)=\frac{k_{J}}{k^{+}}\Delta \widetilde{G}^{I J}(k) \,.
\end{equation}
The equation~\eqref{k_nu-G-tilde-mu-nu} with $\mu = -$ is given by
\begin{equation}
-k^{+}\Delta \widetilde{G}^{- -}(k)+k_{J}\Delta \widetilde{G}^{- J}(k)=0 \,.
\end{equation}
This is satisfied by choosing $\Delta \widetilde{G}^{- -}(k)$ as
\begin{equation}
\Delta \widetilde{G}^{- -}(k)=\frac{k_{J}}{k^{+}}\Delta \widetilde{G}^{- J}(k) \,.
\end{equation}
We have thus learned that we can take $\Delta \widetilde{G}^{I J}(k)$ with $k^2 = 0$
to be independent, and $\Delta \widetilde{G}^{I -}(k)$
and $\Delta \widetilde{G}^{- -}(k)$ are determined
by $\Delta \widetilde{G}^{I J}(k)$.\footnote{
We chose the condition $\Delta \widetilde{G}^{\mu +}(k) = 0$,
and in this gauge choice the transverse field $\Delta \widetilde{G}^{IJ}(k)$
is not traceless and the scalar field corresponding to the dilaton
is contained in the trace part of this transverse field.
This gauge choice simplifies the following analysis
because $\Delta \widetilde{D}(k)$ and $\Delta \widetilde{E}^{\mu}(k)$ vanish.
}

We have solved the first three equations of~\eqref{Psi-tilde-fluctuation-level-(1,1)}.
Let us move on to the remaining two equations:
\begin{eqnarray}
\begin{split}
&\Delta \widetilde{C}(k)+\Delta \widetilde{D}(k)-\sqrt{\frac{\alpha ^{\prime}}{2}}k^{\mu} \Delta \widetilde{F}_{\mu}(k)=0\,,  \\
&\frac{1}{2}\sqrt{\frac{\alpha ^{\prime}}{2}}k^{\mu}\Delta \widetilde{C}(k)  -\frac{\alpha ^{\prime}k^{2}}{4}\Delta \widetilde{F}^{\mu}(k)+\Delta \widetilde{E}^{\mu}(k)-\frac{1}{2}\sqrt{\frac{\alpha ^{\prime}}{2}}k_{\nu}\Delta \widetilde{H}^{\mu \nu}(k)=0 \,.
\end{split}
\label{Psi-tilde-fluctuation-level-(1,1)-O-L-P-V-1}
\end{eqnarray}
We found that $\Delta \widetilde{D}(k)$ and $\Delta \widetilde{E}^{\mu}(k)$ vanish
under the gauge choice~\eqref{G-tilde-gauge}.
Then these equations simplify to
\begin{align}
&\Delta \widetilde{C}(k)-\sqrt{\frac{\alpha ^{\prime}}{2}}k^{\mu} \Delta \widetilde{F}_{\mu}(k)=0\,, \label{Psi-tilde-fluctuation-level-(1,1)-O-L-P-V-2} \\
&\frac{1}{2}\sqrt{\frac{\alpha ^{\prime}}{2}}k^{\mu}\Delta \widetilde{C}(k)  -\frac{\alpha ^{\prime}k^{2}}{4}\Delta \widetilde{F}^{\mu}(k)-\frac{1}{2}\sqrt{\frac{\alpha ^{\prime}}{2}}k_{\nu}\Delta \widetilde{H}^{\mu \nu}(k)=0 \,.\label{Psi-tilde-fluctuation-level-(1,1)-O-L-P-V-3}
\end{align}
We solve~\eqref{Psi-tilde-fluctuation-level-(1,1)-O-L-P-V-2}
for $\Delta \widetilde{C}(k)$ as
\begin{equation}
\Delta \widetilde{C}(k)
= \sqrt{\frac{\alpha ^{\prime}}{2}}k^{\mu} \Delta \widetilde{F}_{\mu}(k) \,,
\end{equation}
and we eliminate $\Delta \widetilde{C}(k)$ from~\eqref{Psi-tilde-fluctuation-level-(1,1)-O-L-P-V-3}
to obtain
\begin{equation}
\frac{\alpha ^{\prime}}{4}k^{\mu} k^{\nu} \Delta \widetilde{F}_{\nu}(k)
-\frac{\alpha ^{\prime}k^{2}}{4}\Delta \widetilde{F}^{\mu}(k)-\frac{1}{2}\sqrt{\frac{\alpha ^{\prime}}{2}}k_{\nu}\Delta \widetilde{H}^{\mu \nu}(k)=0 \,.
\label{F-tilde-H-tilde}
\end{equation}
The relevant gauge transformations are
\begin{eqnarray}
\begin{split}
\delta _{\Lambda} \Delta\widetilde{F}_{\mu}(k)&=\frac{1}{2}\sqrt{\frac{\alpha ^{\prime}}{2}}k_{\mu}\widetilde{\eta}(k)-\widetilde{\zeta}_{\mu}(k) -\frac{1}{2}\sqrt{\frac{\alpha ^{\prime}}{2}}k^{\nu} \widetilde{\omega}_{\mu \nu}(k) \,, \\
\delta _{\Lambda} \Delta\widetilde{H}_{\mu \nu}(k)&=-\sqrt{\frac{\alpha ^{\prime}}{2}}k_{\mu} \widetilde{\zeta}_{\nu}(k) +\sqrt{\frac{\alpha ^{\prime}}{2}}k_{\nu} \widetilde{\zeta}_{\mu}(k) +\frac{\alpha ^{\prime}k^{2}}{4}\widetilde{\omega}_{\mu \nu}(k) \, .
\end{split}
\end{eqnarray}

First consider the case where $k^2 \ne 0$.
In this case, we can impose the condition
\begin{equation}
\Delta \widetilde{H}^{\mu\nu}(k) = 0 \,,
\label{H-tilde=0}
\end{equation}
which can be satisfied
by the gauge transformation using the parameter $\widetilde{\omega}_{\mu \nu}(k)$.
Under this gauge choice, the equation~\eqref{F-tilde-H-tilde} reduces to
\begin{equation}
k^{\mu} k^{\nu} \Delta \widetilde{F}_{\nu}(k)
-k^2 \Delta \widetilde{F}^{\mu}(k)=0 \,.
\label{F-tilde}
\end{equation}
This is nothing but the equation of motion for a free Maxwell gauge field,
and there are no nontrivial solutions when $k^2 \ne 0$.
In fact, we can further impose the condition
\begin{equation}
\Delta \widetilde{F}^{+}(k) = 0 \,,
\end{equation}
which can be satisfied
by the gauge transformation using the parameter $\widetilde{\eta}(k)$
without violating the condition~\eqref{H-tilde=0},
and then the equation~\eqref{F-tilde} with $\mu=+$ is
\begin{equation}
k^+ k^{\nu} \Delta \widetilde{F}_{\nu}(k) = 0 \,.
\end{equation}
Since $k^+ \ne 0$, we obtain
\begin{equation}
k^{\nu} \Delta \widetilde{F}_{\nu}(k) = 0 \,.
\end{equation}
The equation~\eqref{F-tilde} is now given by
\begin{equation}
k^2 \Delta \widetilde{F}^{\mu}(k) = 0 \,,
\end{equation}
and we conclude $\Delta \widetilde{F}^{\mu}(k)=0$ when $k^2 \ne 0$.

Next consider the case where $k^2 = 0$.
In this case we can impose the condition
\begin{equation}
\Delta \widetilde{F}^{\mu}(k) = 0 \,,
\label{F-tilde=0}
\end{equation}
which can be satisfied
by the gauge transformation using the parameter $\widetilde{\omega}_{\mu +}(k)$.
Under this gauge choice, the equation~\eqref{F-tilde-H-tilde} reduces to
\begin{equation}
k_{\nu}\Delta \widetilde{H}^{\mu \nu}(k) = 0 \,.
\label{k_nu-H-tilde-mu-nu}
\end{equation}
We can further impose the condition
\begin{equation}
\Delta \widetilde{H}^{\mu +}(k) = 0 \,,
\label{H-tilde-gauge}
\end{equation}
which can be satisfied
by the gauge transformation using the parameter $\widetilde{\zeta}_{\mu}(k)$
and we adjust the parameter $\widetilde{\omega}_{\mu +}(k)$
so that this condition is compatible with the condition~\eqref{F-tilde=0}.
Under the condition~\eqref{H-tilde-gauge},
the remaining components of $\Delta \widetilde{H}^{\mu \nu}(k)$ are
\begin{equation}
(\Delta \widetilde{H}^{IJ}(k), \, \Delta \widetilde{H}^{I-}(k) \, )\,.
\end{equation}
The equation~\eqref{k_nu-H-tilde-mu-nu} with $\mu = +$
is satisfied by our gauge choice~\eqref{H-tilde-gauge}.
The equation~\eqref{k_nu-H-tilde-mu-nu} with $\mu = I$ is given by
\begin{equation}
-k^{+}\Delta \widetilde{H}^{I -}(k)+k_{J}\Delta \widetilde{H}^{I J}(k) = 0 \,.
\end{equation}
This is satisfied by choosing $\Delta \widetilde{H}^{I -}(k)$ as
\begin{equation}
\Delta \widetilde{H}^{I -}(k)=\frac{k_{J}}{k^{+}}\Delta \widetilde{H}^{I J}(k) \,.
\end{equation}
Then the equation~\eqref{k_nu-H-tilde-mu-nu} with $\mu = -$ is also solved
because
\begin{equation}
k_{I}\Delta \widetilde{H}^{- I}(k)
= {}-\frac{k_{I} k_{J}}{k^{+}}\Delta \widetilde{H}^{I J}(k) = 0 \,,
\end{equation}
where we used $\Delta \widetilde{H}^{\nu \mu}(k) = {}-\Delta \widetilde{H}^{\mu  \nu}(k)$.
We have thus learned that we can take $\Delta \widetilde{H}^{I J}(k)$ with $k^2 = 0$
to be independent, and $\Delta \widetilde{H}^{I -}(k)$
is determined by $\Delta \widetilde{H}^{I J}(k)$.

To summarize, we conclude that all the solutions to~\eqref{Psi-tilde-fluctuation-level-(1,1)}
are equivalent under the gauge transformations
to the free massless propagation of transverse components of
the symmetric tensor field $\Delta \widetilde{G}^{I J}(k)$
and transverse components of the antisymmetric tensor field $\Delta \widetilde{H}^{I J}(k)$.
These are the extra free fields from
the string field $\widetilde{\Psi}_{(1,1)}^{\mathrm{even}}$.\footnote{
In the context of the BRST cohomology on states of ghost number $3$
at the level $(1,1)$,
physical states can be represented
by $\alpha ^{I}_{-1} \widetilde{\alpha} ^{J}_{-1} c_{0} c_{1} \widetilde{c}_{1} \ket{0;k}$
and $\alpha ^{I}_{-1} \widetilde{\alpha} ^{J}_{-1} \widetilde{c}_{0} c_{1} \widetilde{c}_{1} \ket{0;k}$ with $k^2 = 0$,
and we have two copies
of the dilaton, the graviton, and the antisymmetric tensor field.
When we decompose the states based on the world-sheet parity,
one copy is from $\widetilde{\Psi}_{(1,1)}^{\mathrm{even}}$
and the other is from $\widetilde{\Psi}_{(1,1)}^{\mathrm{odd}}$,
in accordance with our conclusion.
}

\subsubsection{Confirmation of no additional conditions}
\label{subsection-3.3.4}
We finally show that the equations
\begin{equation}
\begin{split}
&\frac{\alpha ^{\prime}k^{2}}{4}\widetilde{D}_{\ast}(k) -\sqrt{\frac{\alpha ^{\prime}}{2}}k_{\mu} \widetilde{E}^{\mu}_{\ast}(k)=\mathcal{J}_{D}(k), \\
&\frac{1}{2}\sqrt{\frac{\alpha ^{\prime}}{2}}k^{\mu} \widetilde{D}_{\ast}(k)+ \widetilde{E}^{\mu}_{\ast}(k)+\frac{1}{2}\sqrt{\frac{\alpha ^{\prime}}{2}}k_{\nu} \widetilde{G}^{\mu \nu}_{\ast}(k) =\mathcal{J}_{E}^{\mu}(k)\, , \\
&\frac{1}{4}\sqrt{\frac{\alpha ^{\prime}}{2}}\bigg( k^{\nu}\widetilde{E}^{\mu}_{\ast}(k)+k^{\mu}\widetilde{E}^{\nu}_{\ast}(k)\bigg)+\frac{\alpha ^{\prime}k^{2}}{16}\widetilde{G}^{\mu \nu}_{\ast}(k)=\mathcal{J}_{G}^{\mu \nu}(k)\, ,\\
&\widetilde{C}_{\ast}(k)+\widetilde{D}_{\ast}(k)-\sqrt{\frac{\alpha ^{\prime}}{2}}k_{\mu}\widetilde{F}^{\mu}_{\ast}(k)=\mathcal{J}_{B}(k)\,,  \\
&\frac{1}{2}\sqrt{\frac{\alpha ^{\prime}}{2}}k^{\mu} \widetilde{C}_{\ast}(k)  -\frac{\alpha ^{\prime}k^{2}}{4}\widetilde{F}^{\mu}_{\ast}(k)+ \widetilde{E}^{\mu}_{\ast}(k)-\frac{1}{2}\sqrt{\frac{\alpha ^{\prime}}{2}}k_{\nu} \widetilde{H}^{\mu \nu}_{\ast}(k)=\mathcal{J}_{A}^{\mu}(k) \,, 
\end{split}
\label{Psi-tilde-component-equation-level-(1,1)}
\end{equation}
can be solved for $\widetilde{C}_{\ast}(k)$, $\widetilde{D}_{\ast}(k)$, $\widetilde{E}^{\mu}_{\ast}(k)$, $\widetilde{F}^{\mu}_{\ast}(k)$, $ \widetilde{G}^{\mu \nu}_{\ast}(k)$, and $\widetilde{H}^{\mu \nu}_{\ast}(k)$ when the source terms $\mathcal{J}_{B}(k)$, $\mathcal{J}_{D}(k)$, $\mathcal{J}_{A}^{\mu}(k)$, $\mathcal{J}_{E}^{\mu}(k)$, and $\mathcal{J}_{G}^{\mu \nu}(k)$
are constructed from $\Psi$ satisfying
\begin{equation}
Q_{B}\Psi+\sum_{n=2}^{\infty}\frac{g^{n-1}}{n!} B \, [ \, \Psi ^{n} \, ] = 0 \,.
\label{Psi-equation-level-(1,1)}
\end{equation}
Among the six component fields,
the three fields $\widetilde{C}_{\ast}(k)$, $\widetilde{F}^{\mu}_{\ast}(k)$, and $\widetilde{H}^{\mu \nu}_{\ast}(k)$ appear only in the last two equations.
Let us first consider the first three equations
which only contain $\widetilde{D}_{\ast}(k)$, $\widetilde{E}^{\mu}_{\ast}(k)$, and $ \widetilde{G}^{\mu \nu}_{\ast}(k)$:
\begin{equation}
\begin{split}
\frac{\alpha ^{\prime}k^{2}}{4}\widetilde{D}_{\ast}(k) -\sqrt{\frac{\alpha ^{\prime}}{2}}k_{\mu} \widetilde{E}^{\mu}_{\ast}(k)&=\mathcal{J}_{D}(k), \\
\frac{1}{2}\sqrt{\frac{\alpha ^{\prime}}{2}}k^{\mu} \widetilde{D}_{\ast}(k)+ \widetilde{E}^{\mu}_{\ast}(k)+\frac{1}{2}\sqrt{\frac{\alpha ^{\prime}}{2}}k_{\nu} \widetilde{G}^{\mu \nu}_{\ast}(k) &=\mathcal{J}_{E}^{\mu}(k)\, , \\
\frac{1}{4}\sqrt{\frac{\alpha ^{\prime}}{2}}\bigg( k^{\nu}\widetilde{E}^{\mu}_{\ast}(k)+k^{\mu}\widetilde{E}^{\nu}_{\ast}(k)\bigg)+\frac{\alpha ^{\prime}k^{2}}{16}\widetilde{G}^{\mu \nu}_{\ast}(k)&=\mathcal{J}_{G}^{\mu \nu}(k) \,.
\end{split}
\label{Psi-tilde-component-equation-level-(1,1)-first-three}
\end{equation}
They coincide with the first three equations of~\eqref{comp_eq_Psi_(1,1)},
\begin{equation}
\begin{split}
\frac{\alpha ^{\prime}k^{2}}{4}D(k)-\sqrt{\frac{\alpha ^{\prime}}{2}}k^{\mu}E_{\mu}(k) & =\mathcal{J}_{D}(k) \,, \\
\frac{1}{2}\sqrt{\frac{\alpha ^{\prime}}{2}}k^{\mu}D(k)+E^{\mu}(k)+\frac{1}{2}\sqrt{\frac{\alpha ^{\prime}}{2}}k_{\nu}G^{\mu \nu}(k)
& = \mathcal{J}_{E}^{\mu}(k) \,, \\
\frac{1}{4}\sqrt{\frac{\alpha ^{\prime}}{2}}k^{\mu}E^{\nu}(k)  +\frac{1}{4}\sqrt{\frac{\alpha ^{\prime}}{2}}k^{\nu}E^{\mu}(k) +\frac{\alpha ^{\prime}k^{2}}{16}G^{\mu \nu}(k)
& = \mathcal{J}_{G}^{\mu \nu}(k) \,,
\end{split}
\end{equation}
under the replacement
of $\widetilde{D}_{\ast}(k)$, $\widetilde{E}^{\mu}_{\ast}(k)$, and $\widetilde{G}^{\mu \nu}_{\ast}(k)$
by $D_{\ast}(k)$, $E^{\mu}_{\ast}(k)$, and $G^{\mu \nu}_{\ast}(k)$, respectively.
The equations~\eqref{comp_eq_Psi_(1,1)} are 
those of~\eqref{Psi-equation-level-(1,1)} at the level $(1,1)$
so that we conclude that the equations \eqref{Psi-tilde-component-equation-level-(1,1)-first-three} can be solved for $\widetilde{D}_{\ast}(k)$, $\widetilde{E}^{\mu}_{\ast}(k)$, and $\widetilde{G}^{\mu \nu}_{\ast}(k)$
when the equation~\eqref{Psi-equation-level-(1,1)} has a solution.

The remaining equations,
\begin{equation}
\begin{split}
&\widetilde{C}_{\ast}(k)+\widetilde{D}_{\ast}(k)-\sqrt{\frac{\alpha ^{\prime}}{2}}k_{\mu}\widetilde{F}^{\mu}_{\ast}(k)=\mathcal{J}_{B}(k)\,,  \\
&\frac{1}{2}\sqrt{\frac{\alpha ^{\prime}}{2}}k^{\mu} \widetilde{C}_{\ast}(k)  -\frac{\alpha ^{\prime}k^{2}}{4}\widetilde{F}^{\mu}_{\ast}(k)+ \widetilde{E}^{\mu}_{\ast}(k)-\frac{1}{2}\sqrt{\frac{\alpha ^{\prime}}{2}}k_{\nu} \widetilde{H}^{\mu \nu}_{\ast}(k)=\mathcal{J}_{A}^{\mu}(k) \,, 
\end{split}
\label{Psi-tilde-component-equation-level-(1,1)-2}
\end{equation}
are now regarded as the equations
for $\widetilde{C}_{\ast}(k)$, $\widetilde{F}^{\mu}_{\ast}(k) $, and $ \widetilde{H}^{\mu \nu}_{\ast}(k)$
when $\widetilde{D}_{\ast}(k)$, $\widetilde{E}^{\mu}_{\ast}(k)$, and $\widetilde{G}^{\mu \nu}_{\ast}(k)$
satisfying~\eqref{Psi-tilde-component-equation-level-(1,1)-first-three} are given.
We look for a solution in a gauge where $\widetilde{F}^{\mu}_{\ast}(k)$ vanishes:
\begin{equation}
\widetilde{F}^{\mu}_{\ast}(k) =0 \,.
\end{equation}
This condition can be satisfied by the gauge transformation~\eqref{gauge-even-transformation-level(1,1)-g.n.3} with the parameter~$\widetilde{\zeta}_{\mu}(k)$.
Under this gauge condition, the equations for $\widetilde{C}_{\ast}(k)$ and $\widetilde{H}^{\mu \nu}_{\ast}(k)$ are
\begin{equation}
\begin{split}
&\widetilde{C}_{\ast}(k)+\widetilde{D}_{\ast}(k)=\mathcal{J}_{B}(k)\,,  \\
&\frac{1}{2}\sqrt{\frac{\alpha ^{\prime}}{2}}k^{\mu} \widetilde{C}_{\ast}(k) + \widetilde{E}^{\mu}_{\ast}(k)-\frac{1}{2}\sqrt{\frac{\alpha ^{\prime}}{2}}k_{\nu} \widetilde{H}^{\mu \nu}_{\ast}(k)=\mathcal{J}_{A}^{\mu}(k) \,.
\end{split}
\label{Psi-tilde-component-equation-level-(1,1)-3}
\end{equation}
The first equation can be solved for $\widetilde{C}_{\ast}(k)$ as
\begin{equation}
\widetilde{C}_{\ast}(k)=-\widetilde{D}_{\ast}(k)+\mathcal{J}_{B}(k) \,,
\label{Psi-tilde-component-solution-level-(1,1)-1}
\end{equation}
and we eliminate $\widetilde{C}_{\ast}(k)$ from the second equation to obtain
\begin{equation}
{}-\frac{1}{2}\sqrt{\frac{\alpha ^{\prime}}{2}}k^{\mu}
\widetilde{D}_{\ast}(k)
+\frac{1}{2}\sqrt{\frac{\alpha ^{\prime}}{2}}k^{\mu} \mathcal{J}_{B}(k)
+ \widetilde{E}^{\mu}_{\ast}(k)-\frac{1}{2}\sqrt{\frac{\alpha ^{\prime}}{2}}k_{\nu} \widetilde{H}^{\mu \nu}_{\ast}(k)=\mathcal{J}_{A}^{\mu}(k) \,.
\end{equation}
Thus the equation we need to solve
for the antisymmetric tensor field $\widetilde{H}^{\mu \nu}_{\ast}(k)$ is
\begin{equation}
k_{\nu} \widetilde{H}^{\mu \nu}_{\ast}(k)=\mathcal{J}^{\mu}(k) \,,
\label{H-tilde-source}
\end{equation}
where
\begin{equation}
\mathcal{J}^{\mu}(k)
= {}-k^{\mu} \widetilde{D}_{\ast}(k) +k^{\mu} \mathcal{J}_{B}(k)
+2 \sqrt{\frac{2}{\alpha ^{\prime}}}\widetilde{E}^{\mu}_{\ast}(k)-2 \sqrt{\frac{2}{\alpha ^{\prime}}}\mathcal{J}_{A}^{\mu}(k) \, .
\end{equation}
By contracting~\eqref{H-tilde-source} with $k_\mu$,
we find that \eqref{H-tilde-source} does not allow any solutions
when $k_\mu \mathcal{J}^{\mu}(k)$ is nonvanishing.
We can show that $k_\mu \mathcal{J}^{\mu}(k)$ vanishes in the following way.
We first express $k_\mu \mathcal{J}^{\mu}(k)$ in terms of the source terms as 
\begin{equation}
\begin{split}
k_{\mu} \mathcal{J}^{\mu}(k)
& = {}-k^2 \widetilde{D}_{\ast}(k) +k^2 \mathcal{J}_{B}(k)
+2 \sqrt{\frac{2}{\alpha ^{\prime}}} \, k_\mu \, \widetilde{E}^{\mu}_{\ast}(k)-
2 \sqrt{\frac{2}{\alpha ^{\prime}}} \, k_\mu \mathcal{J}_{A}^{\mu}(k) \\
& = {}-\frac{4}{\alpha ^{\prime}} \, \mathcal{J}_{D}(k)
+k^2 \mathcal{J}_{B}(k)
-2 \sqrt{\frac{2}{\alpha ^{\prime}}} \, k_\mu \mathcal{J}_{A}^{\mu}(k) \,,
\end{split}
\end{equation}
where we used the first equation of~\eqref{Psi-tilde-component-equation-level-(1,1)-first-three}.
The equations~\eqref{Psi-tilde-component-equation-level-(1,1)} correspond to
\begin{equation}
Q_{B}\widetilde{\Psi} = -\sum_{n=2}^{\infty}\frac{g^{n-1}}{n!}[ \, \Psi ^{n} \, ]
\end{equation}
at the level~$(1,1)$,
and we can show that the right-hand side of this equation
is annihilated by the BRST operator
using the equation~\eqref{Psi-equation-level-(1,1)} and the relations in~\eqref{h}.
At the level~$(1,1)$, this yields the following relations
among the source terms:
\begin{equation}
\begin{split}
\frac{1}{2}\sqrt{\frac{\alpha ^{\prime}}{2}}k^{\mu}\mathcal{J}_{D}(k)-\frac{\alpha ^{\prime}k^{2}}{4}\mathcal{J}_{E}^{\mu}(k)+k_{\nu}(\mathcal{J}_{G}^{\mu \nu}(k)+\mathcal{J}_{G}^{\nu \mu}(k))&=0\,, \\
\mathcal{J}_{D}(k)-\frac{\alpha ^{\prime}k^{2}}{4}\mathcal{J}_{B}(k)+\sqrt{\frac{\alpha ^{\prime}}{2}}k_{\mu}\mathcal{J}_{A}^{\mu}(k)&=0\,.
\end{split}
\label{source-equation-level-(1,1)}
\end{equation}
It follows from the second relation that $k_\mu \mathcal{J}^{\mu}(k)$ vanishes
when the source terms are constructed from $\Psi$ satisfying~\eqref{Psi-equation-level-(1,1)}.

The equation we need to solve for $\widetilde{H}^{\mu \nu}_{\ast}(k)$ is
\begin{equation}
k_{\nu} \widetilde{H}^{\mu \nu}_{\ast}(k)=\mathcal{J}^{\mu}(k)
\label{V-equation}
\end{equation}
with $\mathcal{J}^{\mu}(k)$ satisfying
\begin{equation}
k_{\mu}\mathcal{J}^{\mu}(k) = 0 \,.
\end{equation}
In the coordinate basis, the equation is expressed as
\begin{equation}
\partial _{\nu} \widetilde{H}^{\mu \nu}_{\ast}(x^0, x^1, \ldots , x^{25})=\mathcal{J}^{\mu}(x^0, x^1, \ldots , x^{25})
\end{equation}
with $\mathcal{J}^{\mu}(x^0, x^1, \ldots , x^{25})$ satisfying
\begin{equation}
\partial _{\mu} \mathcal{J}^{\mu}(x^0, x^1, \ldots , x^{25}) = 0 \,.
\end{equation}
We look for a solution where only $\widetilde{H}^{\mu 1}_{\ast}(x^0, x^1, \ldots , x^{25})$
and $\widetilde{H}^{1 \mu}_{\ast}(x^0, x^1, \ldots , x^{25})$ are nonvanishing.
In this case we have
\begin{equation}
\frac{\partial}{\partial x^1} \widetilde{H}^{\mu 1}_{\ast}(x^0, x^1, \ldots , x^{25})=\mathcal{J}^{\mu} (x^0, x^1, \ldots , x^{25}) \quad
\text{for} \quad \mu \ne 1 \,.
\end{equation}
The components $\widetilde{H}^{\mu 1}_{\ast}(x^0, x^1, \ldots , x^{25})$
of the antisymmetric tensor field are therefore given by
\begin{equation}
\widetilde{H}^{\mu 1}_{\ast}(x^0, x^1, \ldots , x^{25})
= \int_{-\infty}^{x^1} dx' \mathcal{J}^{\mu} (x^0, x', x^2, \ldots , x^{25}) 
\quad \text{for} \quad \mu \ne 1 \,.
\end{equation}
Then the components $\widetilde{H}^{1 \mu}_{\ast}(x^0, x^1, \ldots , x^{25})$ are
\begin{equation}
\widetilde{H}^{1 \mu}_{\ast}(x^0, x^1, \ldots , x^{25})
= {}-\int_{-\infty}^{x^1} dx' \mathcal{J}^{\mu} (x^0, x', x^2, \ldots , x^{25}) 
\quad \text{for} \quad \mu \ne 1 \,.
\end{equation}
The remaining equation,
\begin{equation}
\partial _{\nu} \widetilde{H}^{1 \nu}_{\ast}(x^0, x^1, \ldots , x^{25})
=\mathcal{J}^{1}(x^0, x^1, \ldots , x^{25}) \,,
\end{equation}
is also solved because
\begin{equation}
\begin{split}
\partial _{\nu} \widetilde{H}^{1 \nu}_{\ast}(x^0, x^1, \ldots , x^{25})
& = {}-\int_{-\infty}^{x^1} dx' \frac{\partial}{\partial x^0} \mathcal{J}^{0} (x^0, x', x^2, \ldots , x^{25}) \\
& \quad~ {}-\int_{-\infty}^{x^1} dx' \frac{\partial}{\partial x^I} \mathcal{J}^{I} (x^0, x', x^2, \ldots , x^{25}) \\
& = \int_{-\infty}^{x^1} dx' \frac{\partial}{\partial x'} \mathcal{J}^{1} (x^0, x', x^2, \ldots , x^{25}) \\
& = \mathcal{J}^{1} (x^0, x^1, x^2, \ldots , x^{25}) \,,
\end{split}
\end{equation}
where we used $\partial _{\mu}\mathcal{J}^{\mu}(x^0, x^1, \ldots , x^{25})=0$.

To summarize, we have shown that the equations~\eqref{Psi-tilde-component-equation-level-(1,1)} can be solved for $\widetilde{C}_{\ast}(k)$, $\widetilde{D}_{\ast}(k)$, $\widetilde{E}^{\mu}_{\ast}(k)$, $\widetilde{F}^{\mu}_{\ast}(k)$, $ \widetilde{G}^{\mu \nu}_{\ast}(k)$, and $\widetilde{H}^{\mu \nu}_{\ast}(k)$ when the source terms $\mathcal{J}_{B}(k)$, $\mathcal{J}_{D}(k)$, $\mathcal{J}_{A}^{\mu}(k)$, $\mathcal{J}_{E}^{\mu}(k)$, and $\mathcal{J}_{G}^{\mu \nu}(k)$
are constructed from $\Psi$ satisfying~\eqref{Psi-equation-level-(1,1)}.
This means that the equations~\eqref{Psi-tilde-component-equation-level-(1,1)}
do not impose any additional conditions
on $\Psi$ satisfying~\eqref{Psi-equation-level-(1,1)}.
This completes the demonstration
that the equations of motion~\eqref{comp_eom_psi_tilde}
and~\eqref{comp_eq_Psi_tilde_(1,1)}
of closed string field theory without the level-matching condition
are equivalent
to the equation of motion~\eqref{comp_eq_Psi_(1,1)_level-matching}
of closed string field theory with the level-matching condition
up to extra free fields described by $\Delta \widetilde{G}^{I J}(k)$
and $\Delta \widetilde{H}^{I J}(k)$.

\section{Conclusions and discussion}
\label{conclusions-discussion}
\setcounter{equation}{0}

We constructed closed bosonic string field theory
without imposing the constraints
\begin{equation}
L_0^- \Psi = 0 \,, \qquad b_0^- \Psi = 0
\end{equation}
on the closed string field $\Psi$.
This is the first implementation of general covariance
in the context of string theory
without using the level-matching condition.
In the free theory, the spacetime metric is represented
by the component field $G_{\mu \nu}(k)$ in~\eqref{Psi_(1,1)-expansion} from $\Psi$,
and the general coordinate transformation is parameterized
by $\xi_{\mu}(k)$ in~\eqref{Lambda_(1,1)-expansion} from $\Lambda$.
In the interacting theory,
not only $\Psi$ but also $\widetilde{\Psi}$ transforms
under the gauge transformation with the parameter $\Lambda$,
as can be seen from~\eqref{m}.
While $\widetilde{\Psi}$ describes the extra free fields,
it does transform under the general coordinate transformation.
Even in closed string field theory with the level-matching condition,
the general covariance is not implemented
by simply replacing ordinary derivatives with covariant derivatives,
but in closed string field theory without the level-matching condition
its implementation is more exotic.
This is why the extra free fields from $\widetilde{\Psi}$
do not couple to gravity despite the fact that they have kinetic terms.\footnote{
This was also the case for the theory constructed by Sen in~\cite{Sen:2015uaa},
where the level-matching condition was imposed on the closed string field
but extra string fields were introduced for the covariant treatment of the Ramond sector.
}

We have demonstrated
in perturbation theory with respect to the string coupling constant~$g$
that closed bosonic string field theory without the level-matching condition
is equivalent
to closed bosonic string field theory with the level-matching condition.
Nonperturbatively, however, the two theories can be inequivalent.
As we commented in subsection~\ref{subsection-3.3.2},
the component field $A_\mu (k)$ in~\eqref{Psi_(1,1)-expansion} vanishes
up to the gauge transformations
in any solutions to the equations of motion,
but we can have solutions which cannot be brought
to the form where $A^{\mu}(k)=0$ by the gauge transformation
if we compactify the target space on a torus.
We expect that there will be a lot of such nonperturbative solutions
for generic backgrounds.
It will be also possible
that there are similar nonperturbative solutions
in superstring field theory
without any constraints for the Ramond sector~\cite{Sen:2015uaa}
and consequently the theory can be inequivalent
to superstring field theory
with the constraint
for the Ramond sector~\cite{Kunitomo:2015usa, Erler:2016ybs, Konopka:2016grr}.
It would be interesting to explore
more about such nonperturbative differences.

In the formulations of open superstring field theory based on the constraint
for the Ramond sector~\cite{Kunitomo:2015usa, Erler:2016ybs, Konopka:2016grr},
the operator $X$~\eqref{X} plays a distinctive role,
and it seems difficult to replace it
with a different operator such as the zero mode of the picture-changing operator.
On the other hand,
the operator $X$ does not have a special meaning
in the approach by Sen~\cite{Sen:2015uaa},
and in fact the zero mode of the picture-changing operator
was used instead of $X$.
In closed bosonic string field theory
with the level-matching condition,
the operator $B$~\eqref{B-introduction} plays a distinctive role,
and it has been difficult to replace it
with a different operator.
On the other hand,
the operator $B$ does not have a special meaning
in closed bosonic string field theory
without the level-matching condition,
and we expect that it is possible to replace it with a different operator.
For example, a family of gauges called linear $b$-gauges
were introduced for open bosonic string field theory in~\cite{Kiermaier:2007jg},
and the operator $b_0$ in Siegel gauge was replaced
with a more general operator made of the $b$ ghost.
The corresponding generalization may be possible
for closed bosonic string field theory
without the level-matching condition,
where $b_0 -\widetilde{b}_0$ is replaced with a more general operator
made of the $b$ ghost and the $\widetilde{b}$ ghost
and $L_0 -\widetilde{L}_0$ is also replaced correspondingly.
In particular, Schnabl gauge can be realized
in a singular limit of linear $b$-gauges,
and it would be interesting to consider the corresponding limit
in closed bosonic string field theory
without the level-matching condition.
More exotic choices might also be possible
and, for example, it would be interesting
to consider choosing $b_0$ in place of $b_0 -\widetilde{b}_0$
and $L_0$ in place of $L_0 -\widetilde{L}_0$.
Furthermore, there might be more general constructions
of multi-string products which satisfy~\eqref{h}.
While we have not understood the reason clearly,
somehow use of extra free string fields
seems to make string field theory more flexible.

This flexibility might play a role
when we try to extract closed strings from open strings
in the context of the AdS/CFT correspondence.
For example, the world-sheet of closed strings
is constructed from the world-sheet of open strings
via unconventional gluing
in the hexagon approach~\cite{Basso:2015zoa},
and the representation of closed strings
without using the level-matching condition
might be useful.

We usually use the Batalin-Vilkovisky formalism
for gauge fixing of string field theory,
but the construction of a classical master action
in the Batalin-Vilkovisky formalism can be complicated.
For closed bosonic string field theory
with the level-matching condition,
the construction of a classical master action is straightforward
because the action~\eqref{a}
and the gauge transformation~\eqref{i}
are both written in terms of the same set of multi-string products
satisfying the $L_\infty$ relations~\eqref{g}.
The corresponding relations~\eqref{h}
for closed bosonic string field theory
without the level-matching condition
are different from $L_\infty$ relations,
but they are closely related to $L_\infty$ relations
and we expect that a classical master action
can be easily constructed
following the analogous construction in~\cite{Sen:2015uaa}.

The approach by Sen~\cite{Sen:2015uaa} to the covariant treatment of the Ramond sector
using spurious free fields
has opened a new direction of research in string field theory,
and we believe that we have revealed that the approach has a counterpart
which is related to the level-matching condition on closed string fields.
It will be also possible
to combine the two
and formulate closed superstring field theory
without imposing the level-matching condition
and the constraint on the Ramond sector.
Some aspects of the new approach are still mysterious,
and we hope that the results of this paper will help demystify
the potential of this interesting new direction of research in string field theory.
 
\bigskip
\noindent
{\bf \large Acknowledgments}

\medskip
The results of this paper were presented at the {\it Discussion Meeting on String Field Theory and String Phenomenology} held at the Harish-Chandra Research Institute in February of 2018, and we thank the audience of the meeting for valuable discussions especially on an issue in our previous construction of the interactions. In particular, we are grateful to Ted Erler and Ashoke Sen for detailed discussions.
We also thank Harold Erbin for recent useful discussions. 
The work of Y.O. was supported in part by a Grant-in-Aid for Scientific Research~(C)~17K05408
from the Japan Society for the Promotion of Science (JSPS).

\end{document}